\documentclass[journal]{IEEEtran}

\usepackage{algorithm, algorithmic}

\usepackage{subfigure}
\usepackage{textcomp}
\usepackage{stfloats}
\usepackage{url}
\usepackage{verbatim}
\usepackage{hyperref}					
\hypersetup{colorlinks,
	linkcolor=blue,%
	anchorcolor=blue,
    citecolor=blue}

\usepackage{bm}

\renewcommand{\bf}[1]{\bm{#1}}

\usepackage{cite}
\usepackage{amsmath,amssymb,amsfonts}

\usepackage{amsthm}

\theoremstyle{definition}

\newtheorem{theorem}{Theorem}
\newtheorem{corollary}{Corollary}
\newtheorem{remark}{Remark}
\newtheorem{assumption}{Assumption}

\usepackage{algorithmic}
\usepackage{graphicx}

\usepackage{textcomp}
\usepackage{xcolor}

\usepackage{array} 
\renewcommand\arraystretch{1.75}

\usepackage{tabularx}

\usepackage{array} 
\renewcommand\arraystretch{1.1}  

\usepackage{booktabs}

\begin{document}
	
	
	\title{\huge Probabilistic Constellation Shaping for OFDM ISAC \\Signals Under Temporal-Frequency Filtering}
	
	\author{Zhen Du,~\IEEEmembership{Member,~IEEE,} Jingjing Xu, Yifeng Xiong,~\IEEEmembership{Member,~IEEE,} Jie Wang, \\Musa Furkan Keskin,~\IEEEmembership{Member,~IEEE,} Henk Wymeersch,~\IEEEmembership{Fellow,~IEEE},  \\Fan Liu,~\IEEEmembership{Senior Member,~IEEE}, and~Shi Jin,~\IEEEmembership{Fellow,~IEEE}
		\thanks{(\textit{Corresponding author: Fan Liu})}
		\thanks{Zhen Du, Jingjing Xu and Jie Wang are with the School of Electronic and Information Engineering, Nanjing University of Information Science and Technology, Nanjing 210044, China (email: duzhen@nuist.edu.cn).
		}
		\thanks{Yifeng Xiong is with the School of Information and Electronic Engineering, Beijing University of Posts and Telecommunications, Beijing 100876, China (e-mail: yifengxiong@bupt.edu.cn).
		}
		\thanks{Musa Furkan Keskin and Henk Wymeersch are with Department of Electrical Engineering, Chalmers University of Technology, 41296 Gothenburg, Sweden (e-mail: {furkan; henkw}@chalmers.se).
		}		
		\thanks{Fan Liu and Shi Jin are with the National Mobile Communications Research Laboratory, Southeast University, Nanjing 210096, China (e-mail: fan.liu; jinshi@seu.edu.cn).
		}
	}

	\maketitle
	
	\begin{abstract}
		Integrated sensing and communications (ISAC) is considered an innovative technology in sixth-generation (6G) wireless networks, where utilizing orthogonal frequency division multiplexing (OFDM) communication signals for sensing provides a cost-effective solution for implementing ISAC. However, the sensing performance of matched and mismatched filtering schemes can be significantly deteriorated due to the signaling randomness induced by finite-alphabet modulations with non-constant modulus, such as quadrature amplitude modulation (QAM) constellations. Therefore, improving sensing performance without significantly compromising communication capability (i.e., maintaining randomness), remains a challenging task. To that end, we propose a unified probabilistic constellation shaping (PCS) framework that is compatible with both matched and mismatched filtering schemes, by maximizing the communication rate while imposing constraints on mean square error (MSE) of sensing channel state information (CSI), power, and probability distribution. Specifically, the MSE of sensing CSI is leveraged to optimize sensing capability, which is illustrated to be a more comprehensive metric compared to the output SNR after filtering ($\text{SNR}_\text{out}$) and integrated sidelobes ratio (ISLR). Additionally, the internal relationships among these three sensing metrics are explicitly analyzed. Finally, both simulations and field measurements validate the efficiency of proposed PCS approach in achieving a flexible S\&C trade-off, as well as its credibility in enhancing 6G wireless transmission in real-world scenarios.
	\end{abstract}
	

	\section{Introduction}
	Integrated sensing and communications (ISAC) has been envisioned as a ground-breaking technology to support 6G usage scenarios such as smart cities, intelligent transportation, and low-altitude economy, where ultra-reliable and high-precision location-aware services are essential \cite{ITU-R,liu2022integrated,lu2024integrated}. The primary advantage of ISAC stems from the dual design of sensing and communication (S\&C) functionalities, which not only enhances the utilization efficiency of both hardware and spectral resources, but also activates mutual gains between S\&C criteria \cite{du2023towards}.
	
	Toward that end, developing a unified waveform that enables simultaneous information transmission and target sensing becomes crucial. To comprehensively balance S\&C qualities of service (QoS) and computational complexities,  while maintaining compatibility with the existing 5G NR protocol \cite{li5GNR}, the communication-centric signaling design \cite{berger2010signal,sturm2011waveform} is believed as a more economically viable paradigm over sensing-centric \cite{hassanien2015dual} and joint design \cite{liu2018mu} schemes. In this context, directly leveraging existing communication signals, such as orthogonal frequency division multiplexing (OFDM), can facilitate sensing tasks \cite{berger2010signal,sturm2011waveform,tigrek2012ofdm,zheng2017super,rodriguez2023supervised,baudais2023doppler,wang2023waveform,keskin2024fundamental,liu2024ofdm,du2024reshaping}. Despite the fact that OFDM has been widely employed as the default waveform in cellular communications and WiFi \cite{liu2024ofdm}, its sensing capability in 6G networks remains under investigation. Overall, the effectiveness of exploiting OFDM communication signals for sensing is mainly determined by two factors: 1) the filtering schemes employed, and 2) the signaling randomness introduced by communication data symbols.

	First, one may take advantage of the classic matched filtering (MF) technique to extract the range-Doppler information when using OFDM communication signals for sensing purposes \cite{berger2010signal}. According to its definition, MF is optimal in terms of maximizing the output signal-to-noise ratio ($\text{SNR}_\text{out}$), given a specific input SNR ($\text{SNR}_\text{in}$). In this context, the filtering output refers to the target response function in the delay-Doppler domain. Particularly, when MF is used, this response function represents the celebrated Woodward ambiguity function \cite{woodward2014probability}. In order to ameliorate sidelobes of response function, commonly quantified by integrated sidelobes ratio (ISLR) \cite{zhen2021multicarrier} or peak sidelobes ratio (PSLR) \cite{rodriguez2023supervised}, an alternative approach is to utilize mismatched filters (MMF) \cite{sturm2011waveform,baudais2023doppler,keskin2024fundamental,wang2023waveform}. 
	For example, reciprocal filtering (RF) based on zero-forcing principle \cite{sturm2011waveform}, can recover the orthogonality of range-Doppler steering matrix by performing element-wise division operation, thereby eliminating the random symbols from target-related echo components and producing a sinc-like response function. Nevertheless, RF comes with a drawback, i.e., the $\text{SNR}_\text{out}$ loss, due to the amplification of background noise. To combine the advantages of MF and RF (i.e., balancing between $\text{SNR}_\text{out}$ and ISLR/PSLR), the Wiener filter (WF) \cite{baudais2023doppler,keskin2024fundamental} offers a promising solution. The WF, which borrows the principle of linear minimum mean square error (LMMSE) equalization \cite{fazel2008multi} in communication systems, is equivalent to RF in noise-free environments, and approaches MF in low $\text{SNR}_\text{in}$ case \cite{keskin2024fundamental}. In addition, some variants of RF, such as loaded-RF and threshold-RF elaborated in \cite{rodriguez2023supervised}, demonstrate similar performance to WF. However, all of the aforementioned works primarily concentrate on optimizing ISLR/PSLR and $\text{SNR}_\text{out}$, which will be illustrated in this paper insufficient to quantify the sensing performance. More critically, those previous works focus on receiver design/optimization while overlooking transmit optimization, a research gap that we try to fill in this work by controlling and optimizing the randomness of ISAC signals.

	Second, most of filtering schemes prefer a flat signaling spectrum to facilitate high-resolution mainlobe and low sidelobes, such as adopting phase shift keying (PSK)-modulated OFDM signals \cite{tigrek2012ofdm}. In contrast, communication systems typically leverage quadrature amplitude modulation (QAM)-modulated OFDM signals due to the larger Euclidean distance of discrete symbols, thereby enhancing the communication reliability. Consequently, adopting QAM-modulated OFDM communication signals for sensing leads to compromised sensing performance. For instance, this may increase sidelobes of response function for MF (i.e., the ambiguity function), or amplify the background noise when using RF \cite{baudais2023doppler}. The core challenge here is therefore to improve sensing capabilities without significantly degrading communication performance. One promising research line along this topic is inspired by the deterministic-random trade-off observed in ISAC systems \cite{xiong2023fundamental}, which provides the flexibility to balance between S\&C tasks by adjusting the input distribution of the ISAC signal. Specifically, a probabilistic constellation shaping (PCS) approach \cite{du2024reshaping,du2023probabilistic,xu2024ofdm} for OFDM-based ISAC signaling was proposed, through a tailored offline optimization of constellations' input probability distribution, which is independent to both the channel state information (CSI) and the random realization of discrete symbols. Prior to that, PCS has been primarily used for minimizing the gap between the achievable information rate (AIR) and Shannon capacity \cite{bocherer2014probabilistic,schulte2015constant,cho2019probabilistic}. Notably, the PCS approach developed in \cite{du2024reshaping,du2023probabilistic,xu2024ofdm}, which focuses on optimizing the fourth-order moment of constellation amplitudes, is only applicable to MF in noise-free environments, without accounting for the MMF schemes and the noise amplification effect.
		
	Based upon the above reasoning, this article aims to tackle a challenge: how to alleviate the negative effect of OFDM signaling randomness on the sensing performance while preserving its communication capability, in order to meet the compatibility with different filtering schemes, unlike the work in \cite{du2024reshaping} where the PCS approach is limited to MF. Therefore, we aim to perform joint transmit/receive design for random OFDM ISAC signaling.
	For clarity, the main contributions of this paper are summarized as follows.
	\begin{itemize}
		\item We establish the relationship among the MSE of sensing CSI, $\text{SNR}_\text{out}$ and ISLR of the target response function. Specifically, we reveal that the MSE is influenced by several factors, including the potentially amplified noise, the sidelobes of the response function, and the peak energy loss of the response function, which are determined by the signaling randomness and filtering schemes simultaneously. As a result, we propose that MSE serves as a more comprehensive criterion for evaluating sensing performance, in contrast to relying solely on $\text{SNR}_\text{out}$ and ISLR. 
		\item We propose to construct a unified PCS approach by adopting the MSE of sensing CSI and AIR as S\&C performance indicators, respectively. In contrast to the technique in \cite{du2024reshaping}, which is only applicable to MF, we provide compatibility with generic MMF schemes. Benefiting from optimizing the input distribution of constellations, this unified PCS framework allows for a flexible trade-off between S\&C performance, which can be tailored to meet practical QoS requirements. Similar to the work in \cite{du2024reshaping}, our PCS approach can be implemented offline and deployed online, enabling a significant potential in practical 6G usages.
		\item Simulations are presented to demonstrate the superiority of WF over RF and MF, both with and without PCS. Moreover, the capability of proposed PCS approach in achieving a flexible trade-off between S\&C performance, is also illustrated under different filtering schemes. In particular, field experiments are conducted using real-world observation data to validate the practical applicability of PCS, highlighting its credibility for enhancing 6G wireless transmission in real-world scenarios.
	\end{itemize}
	
	The remainder of this paper is organized as follows. Sec. \ref{sec2} introduces the S\&C signal models in the ISAC system. Sec. \ref{sec3} elaborates on the sensing criteria including the MSE of sensing CSI, $\text{SNR}_\text{out}$, and ISLR, and builds up the relationship among them. Sec. \ref{sec4} proposes a unified PCS approach compatible to MF and MMF. Sec. \ref{sec5} provides simulation and experiment results to validate the theoretical analysis of the PCS approach. Finally, Sec. \ref{sec6} concludes the paper.
	
	\textit{Notation:} Throughout the paper, $\bf{A}$, $\bf{a}$, and $a$ denote a matrix, vector, and scalar, respectively; $\vert a\vert$, $\vert \bf{A}\vert$ and $\frac{\bf{B}}{\bf{A}}$ represent the modulus of $a$, the element-wise modulus of $\bf{A}$ and the element-wise division of $\bf{B}$ by $\bf{A}$, respectively;
	$\Re(\cdot)$, $\mathbb{E}(\cdot)$, $\Vert\cdot\Vert_F$, $(\cdot)^{T}$, $(\cdot)^{*}$, $(\cdot)^{H}$, $\odot$ and $\bf{1}_N$ denote the real part of a complex number, expectation, Frobenius norm, transpose, conjugate, Hermitian, Hadamard (element-wise) product, and the identity vector of size $N\times 1$, respectively; $\mathcal{CN}(\mu,\sigma^2)$ denotes a complex Gaussian distribution with mean $\mu$ and variance $\sigma^2$; In addition, $\text{sinc}(x)=\frac{\sin(\pi x)}{\pi x}$, and $\delta(x)$ represents Dirac delta function.

	\section{Signal Model}\label{sec2}
	\subsection{Transmit ISAC Signal Model}
	We consider a monostatic ISAC system employing OFDM signals for S\&C tasks simultaneously. The ISAC OFDM signal consisting of $N$ subcarriers and $M$ symbols, and occupying a bandwidth of $B$ Hz and a symbol duration of $T_p$ seconds, is given by
	\begin{equation}\label{equ1}
		\begin{aligned}
			s(t) = \frac{1}{\sqrt{M}}\sum\nolimits^{M-1}_{m=0}s_m(t-mT_\text{sym}),
		\end{aligned}
	\end{equation}
	where
	\begin{equation}\label{equ2}
		\begin{aligned}
			s_m(t) = \frac{1}{\sqrt{N}}\sum\nolimits^{N-1}_{n=0} x_{n,m} e^{j2\pi n \Delta ft} \text{rect}\left( \frac{t-mT_\text{sym}}{T_\text{sym}} \right).
		\end{aligned}
	\end{equation}
	In (\ref{equ2}), $x_{n,m}$ denotes the transmitted frequency-domain data drawn from a given finite alphabet, e.g., 64-QAM constellation. Note that the input distribution of $x_{n,m}$ may be non-uniform. 
	In addition, $\Delta f = B/N=1/T_p$ represents the subcarrier interval in the frequency domain, and $\text{rect}(t)$ represents the rectangle window, equal to $1$ for $0\leq t\leq 1$, and zero otherwise. Here, $T_\text{sym}=T_p+T_{cp}$, where $T_{cp}$ denotes the length of cyclic prefix (CP). In order to eliminate the inter-symbol-interference (ISI) of received S\&C signals, $T_{cp}$ must be larger than round-trip delay of the farthest target/path. 

	\begin{assumption}
		We assume the same input distribution of discrete constellations across subcarriers and symbols. 
		Therefore, we may omit the subscripts $n$ and $m$ when calculating the expectation with respect to the input distribution for notational clarity, e.g., $\mathbb{E}\left\{\sum_{n,m}|x_{n,m}|^2\right\} \triangleq NM\mathbb{E}\left\{|x|^2\right\}$, where the expectation is conducted in terms of the probability distribution $p(x)$.
	\end{assumption}
	
	\begin{assumption}
		The transmit power of a given constellation such as PSK/QAM, is normalized, i.e. $\mathbb{E}\{|x_{n,m}|^2\}=1$ for $\forall n,m$. The total average transmit energy of $M$ symbols is thus expressed as $\mathbb{E}\{|s(t)|^2\} = \frac{1}{NM}\mathbb{E}\left\{\sum_{n,m}|x_{n,m}|^2\right\}=1$.
	\end{assumption}
	
	Finally, the transmit ISAC signal through the up converter can be formulated as
	\begin{equation}\label{equ3}
		\begin{aligned}
			\Re\{s(t)\exp\left(j2\pi f_ct\right)\},
		\end{aligned}
	\end{equation}
	where $f_c$ is the carrier frequency.

	\subsection{Sensing Signal Model}
	Let us exploit OFDM communication signals to detect $Q$ point targets resolvable in the delay-Doppler (DD) domain. Therefore, the reflected echo is formulated as
	\begin{equation}\label{equ4}
		\begin{aligned}
			y_s(t) = \sum\nolimits^Q_{q=1} \alpha_q s(t-\tau_q)e^{j2\pi \nu_q t} + z(t),
		\end{aligned}
	\end{equation}
	where $\alpha_q$, $\tau_q$, $\nu_q$ and $z(t)$ represent the complex channel gain coefficient, the target delay, the Doppler shift of the $q$th target, and the additive white Gaussian noise (AWGN), respectively. Note that $\alpha_q\sim\mathcal{CN}(0,\sigma^2_{\alpha_q})$ and $z(t)\sim\mathcal{CN}(0,\sigma^2)$. Notice that this OFDM echo model omits the range walking effect during $MT_\text{sym}$ due to $\nu_q\leq\frac{f_c}{BMT_\text{sym}}$ \cite{zhang2020joint,hakobyan2019high}, and the inter-carrier-interference (ICI) effect since all subcarriers of echoes are approximately orthogonal in terms of $\nu_q\leq\frac{1}{10}\Delta f$ \cite{sturm2011waveform}.

	
	After symbol synchronization and removing the CP from each symbol, the observed data matrix may be reformulated as a fast-time/slow-time sampling representation, which is expressed as \cite{keskin2021mimo}
	\begin{equation}\label{equ5}
		\begin{aligned}
			\bf{Y}_s = \bf{H}\odot\bf{X}+\bf{Z},
		\end{aligned}
	\end{equation}
	where $\bf{H}\in\mathbb{C}^{N\times M}$ denotes the sensing CSI matrix, formulated as
	\begin{equation}\label{equ6}
		\begin{aligned}
			\bf{H} = \sum\nolimits^Q_{q=1} \alpha_q\bf{b}(\tau_q)\bf{c}^H(\nu_q),
		\end{aligned}
	\end{equation}
	and $\bf{X}\in\mathbb{C}^{N\times M}$ and $\bf{Z}\in\mathbb{C}^{N\times M}$ represent the random symbol and noise matrices, respectively.
	
	In (\ref{equ6}), $\bf{b}(\tau)$ and $\bf{c}(\nu)$ are frequency-domain and temporal (slow-time) steering vectors, expressed as
		\begin{align*}
			\bf{b}(\tau)=& \left[1,e^{-j2\pi \Delta f \tau},\cdots,e^{-j2\pi (N-1)\Delta f \tau}\right]^T, \\
			\bf{c}(\nu) =& \left[1,e^{-j2\pi f_c T_\text{sym} \nu},\cdots,e^{-j2\pi (M-1)f_c T_\text{sym} \nu}\right]^T.
		\end{align*}
	Notably, the $(n,m)$-th element of $\bf{H}$, denoted as $h_{n,m}$, satisfies $h_{n,m}\sim\mathcal{CN}\left(0,\sum_q\sigma^2_{\alpha_q}\right)$. This is due to the fact that the sum of individual Gaussian variables is still Gaussian, with its mean and variance presented as
	\begin{equation}
		\begin{aligned}
			\mathbb{E}\{h_{n,m}\}=\sum_q\mathbb{E}\{\alpha_q\}e^{-j2\pi n\Delta f \tau_q}e^{j2\pi f_cmT_\text{sym} \nu_q}=0
		\end{aligned}
	\end{equation}	
	and
	\begin{equation}
		\begin{aligned}
			& \mathbb{E}\{|h_{n,m}|^2\} = \sum_{q,q'}\mathbb{E}\{\alpha_q\alpha^*_{q'}\}C_{n,m}
			\\ & = \sum_{q}\mathbb{E}\{|\alpha_q|^2\} + \sum_{q,q'\neq q}\mathbb{E}\{\alpha_q\}\mathbb{E}\{\alpha^*_{q'}\}C_{n,m}  =  \sum_q\sigma^2_{\alpha_q},
		\end{aligned}
	\end{equation}	
	where $C_{n,m}=e^{-j2\pi [n\Delta f (\tau_q-\tau_{q'})-f_cmT_\text{sym} (\nu_q-\nu_{q'})]}$.
	
	Next, the DD profile can be obtained in accordance with
	\begin{equation}\label{equ1x}
		\begin{aligned}
			 |\bf{\Lambda}|^2 = \left|\bf{F}^H_N\hat{\bf{H}}\bf{F}_M\right|^2,
		\end{aligned}
	\end{equation}
	where $\bf{F}_N$ represents the $N$-dimensional discrete Fourier transform (DFT) matrix, satisfying $\bf{F}_N=\frac{1}{\sqrt{N}}(W_N^{n_1n_2})_{n_1,n_2=0,1,\cdots,N-1}$ and $W_N=e^{-j2\pi/N}$. $\bf{F}_M$ has the same definition except its dimension.  In addition, $\hat{\bf{H}}$ denotes the estimated counterpart of $\bf{H}$, i.e. the estimation of sensing CSI, expressed as
	\begin{equation}\label{csi}
		\begin{aligned}
			\hat{\bf{H}} = \bf{Y}_s\odot\bf{G},
		\end{aligned}
	\end{equation}	
	where $\bf{G}$ denotes the temporal-frequency (TF) filtering matrix, which is expressed as
	\begin{equation}\label{g}
		\begin{aligned}
			\bf{G} = 
			\left\{
			\begin{array}{ll}
				\bf{X}^*,   & \text{MF}, \\
				\frac{1}{\bf{X}} ,   & \text{RF}, \\
				\frac{\bf{X}^*}{\left\vert \bf{X} \right\vert^2+\text{SNR}_\text{in}^{-1}}, & \text{WF}
			\end{array}
			\right.
		\end{aligned}
	\end{equation}
	Notably, $\text{SNR}_\text{in}=\sum_q\sigma^2_{\alpha_q}/\sigma^2$ represents the input SNR before filtering.

	For notational convenience, below we denote $\chi_{n,m}=x_{n,m}g_{n,m}$ as
	\begin{equation}\label{equ26}
		\begin{aligned}
			\chi_{n,m} =
			\left\{
			\begin{array}{ll}
				x_{n,m}x^*_{n,m} = \left\vert x_{n,m} \right\vert^2, & \text{MF}, \\
				x_{n,m}\frac{1}{x_{n,m}} = 1 ,   & \text{RF}, \\
				\frac{\left\vert x_{n,m} \right\vert^2}{\left\vert x_{n,m} \right\vert^2+\text{SNR}_\text{in}^{-1}}, & \text{WF}
			\end{array}
			\right.
		\end{aligned}
	\end{equation}
	which can be interpreted as the filtered ``spectrum'', representing the target response function in the TF domain. It is worth mentioning that $\chi_{n,m}$ is a real variable with regard to MF, RF and WF.
	
	\vspace{2mm}

	Since $\bf{\Lambda} = \bf{F}^H_N\hat{\bf{H}}\bf{F}_M$ is random with respect to the input distribution, the complex channel gain, and the noise, we thereby concentrate on the expectation of DD profile, i.e., $\mathbb{E}\{|\bf{\Lambda}|^2\}$. To proceed, we first express the $(k,p)$th entry of $\bf{\Lambda}$ as
	\begin{equation}\label{equ27}
		\begin{aligned}
			\Lambda_{k,p} = & \frac{1}{\sqrt{NM}}\sum_{n,m}\left(\sum_{q}\alpha_q e^{-j2\pi \frac{n\tilde{k}_{q}}{N}}e^{j2\pi \frac{m\tilde{p}_{q}}{M}}\chi_{n,m} + z_{n,m}g_{n,m} \right)\\ & \times e^{j2\pi \frac{nk}{N}}e^{-j2\pi \frac{mp}{M}}
			\\ = & \frac{1}{\sqrt{NM}}\sum_{q}\alpha_q \sum_{n,m} \chi_{n,m} e^{-j2\pi \frac{n\tilde{k}_{q}}{N}}e^{j2\pi \frac{m\tilde{p}_{q}}{M}}  e^{j2\pi \frac{nk}{N}}e^{-j2\pi \frac{mp}{M}} 
			\\ & + \frac{1}{\sqrt{NM}} \sum_{n,m} z_{n,m}g_{n,m} e^{j2\pi \frac{nk}{N}}e^{-j2\pi \frac{mp}{M}} 
			\\ = & \sum_{q}\alpha_q r(k-\tilde{k}_q,p-\tilde{p}_q) + \tilde{z}(k,p),
		\end{aligned}
	\end{equation}
	where $k=0,1,\cdots,N-1$ and $p=0,1,\cdots,M-1$ denote the delay and Doppler indices, respectively, and $\tilde{k}_q=N\Delta f \tau_q $ and $\tilde{p}_q=MT_{sym}f_c \nu_q$. Besides,
	\begin{equation}\label{dft}
		\begin{aligned}
			r(k,p) = \frac{1}{\sqrt{NM}}\sum_{n,m} \chi_{n,m} e^{j2\pi \frac{nk}{N}} e^{-j2\pi \frac{mp}{M}}
		\end{aligned}
	\end{equation} 
	represents the target response function (in the DD domain) characterizing the filtering output when hypothesizing a target at a given DD-bin,  
	which is mathematically the 2D-DFT of $\chi_{n,m}$, and $\tilde{z}(k,p)$ is the output Gaussian noise. 
	
	
	We emphasize that $\Lambda_{k,p}$ in (\ref{equ27}) is formulated as a linear combination of $Q$ delay/Doppler-shifted versions of $r(k,p)$, plus the output noise. To successfully detect the targets, one typically identifies $Q$ peaks in the squared filtering output, i.e. $|\Lambda_{k,p}|^2$, using thresholding decision strategy, such as the cell-average constant false-alarm rate (CA-CFAR) algorithm \cite{richards2005fundamentals}. 
	
	\begin{figure*}[!t]
		\begin{equation}\label{eq2}
			\begin{aligned}
				\mathbb{E}\{|\Lambda_{k,p}|^2\} \overset{(a)}{=} & \sigma^2_{\alpha}\frac{1}{NM}\sum_{n,m}\sum_{n',m'}\mathbb{E}\{\chi_{n,m}\chi^*_{n',m'} \} e^{j2\pi \frac{(n-n')(k-\tilde{k})}{N}}e^{-j2\pi \frac{(m-m')(p-\tilde{p})}{M}} \\ & + \frac{1}{NM}\sum_{n,m}\sum_{n',m'}\mathbb{E}\{z_{n,m}g_{n,m}z^*_{n',m'}g^*_{n',m'} \} e^{j2\pi \frac{(n-n')k}{N}}e^{-j2\pi \frac{(m-m')p}{M}}  
				\\ \overset{(b)}{=} & \sigma^2_{\alpha}\frac{1}{NM}\sum_{n,m}\mathbb{E}\{\chi_{n,m}^2 \} + 	\sigma^2_{\alpha}\frac{1}{NM}\sum_{n,m}\sum_{n',m'}\mathbb{E}\{\chi_{n,m}\}\mathbb{E}\{\chi_{n',m'}^*\} e^{j2\pi \frac{(n-n')(k-\tilde{k})}{N}}e^{-j2\pi \frac{(m-m')(p-\tilde{p})}{M}} \\ & - \sigma^2_{\alpha}\frac{1}{NM}\sum_{n,m}\mathbb{E}^2\{\chi_{n,m}\}  + \sigma^2\frac{1}{NM}\sum_{n,m}\mathbb{E}\{|g_{n,m}|^2\} 
				\\ \overset{(c)}{=} & \sigma^2_{\alpha}\underbrace{\left(\mathbb{E}\{\chi^2 \} - \mathbb{E}^2\{\chi\}\right)}_{\text{Var}\left(\chi\right)} + NM\sigma^2_{\alpha}\mathbb{E}^2\{\chi\}\text{sinc}^2\left(k-\tilde{k}\right)\text{sinc}^2\left(p-\tilde{p}\right) + \sigma^2\mathbb{E}\{|g|^2\}
			\end{aligned}
		\end{equation}
		\rule{18cm}{1.0pt}
	\end{figure*}

	In practice, high sidelobe level of strong targets together with amplified noise levels, would incur severe miss detection of weak targets, where a desirable $r(k,p)$ should exhibit low sidelobes. In other words, we only need to concentrate on the behavior of DD response function belonging to \textit{a single target} (with its DD indices as $(\tilde{k},\tilde{p})$) and the noise effect, for simplicity of remaining sensing performance analysis. Specifically, the filtering result is shown in (\ref{eq2}) at the top of this page, which is compatible with all three filtering schemes as well as both uniform and non-uniform constellations. Note that ``$\overset{(a)}{=}$'' holds because the cross-terms cancel out due to the statistical independence between transmit symbols and the receiver noise. Furthermore, ``$\overset{(b)}{=}$'' is valid because the difference between the second and third terms corresponds to a sum over $NM(NM-1)$ components when $n\neq n'$ and/or $m\neq m'$. Finally, ``$\overset{(c)}{=}$'' follows from straightforward algebraic simplification. Notably, Dirichlet kernels are used in (\ref{eq2}), expressed as
	\begin{equation}
		\begin{aligned}
			\sum\nolimits_{n,n'} & e^{j2\pi\frac{(n-n')(k-\tilde{k})}{N}} \notag =  \frac{\sin^2(\pi (k-\tilde{k}))}{\sin^2(\frac{\pi (k-\tilde{k})}{N})}\approx N^2\text{sinc}^2(k-\tilde{k})
			\\
			\sum\nolimits_{m,m'} & e^{j2\pi\frac{(m-m')(p-\tilde{p})}{M}} \notag = \frac{\sin^2(\pi (p-\tilde{p}))}{\sin^2(\frac{\pi (p-\tilde{p})}{M})} \approx M^2\text{sinc}^2(p-\tilde{p})
		\end{aligned}
	\end{equation}
	which can be further simplified as $N^2\delta^2(k-\tilde{k})$ and $M^2\delta^2(p-\tilde{p})$ when DD indices $\tilde{k}$ and $\tilde{p}$ are integers (on-grid targets).

	Since normalized delay-Doppler profiles are typically analyzed, the sensing performance can be quantified using the dynamic range (DR) \cite{hakobyan2017novel}, which is defined as
		\begin{equation}\label{equ160}
			\begin{aligned}
				\text{DR} = &  \frac{\mathbb{E}\{|\Lambda_{\tilde{k},\tilde{p}}|^2\}}{\mathbb{E}\{|\Lambda_{k,p}|^2\}} \approx  \frac{NM\sigma^2_{\alpha}\mathbb{E}^2\{\chi\}+\sigma^2_{\alpha}\text{Var}\left(\chi\right)+\sigma^2\mathbb{E}\{|g|^2\}}{\sigma^2_{\alpha}\text{Var}\left(\chi\right)+\sigma^2\mathbb{E}\{|g|^2\}} \\
				= & 1+  \frac{NM\sigma^2_{\alpha}\mathbb{E}^2\{\chi\}}{\sigma^2_{\alpha}\text{Var}\left(\chi\right)+\sigma^2\mathbb{E}\{|g|^2\}} 
				\approx  \frac{NM\sigma^2_{\alpha}\mathbb{E}^2\{\chi\}}{\sigma^2_{\alpha}\text{Var}\left(\chi\right)+\sigma^2\mathbb{E}\{|g|^2\}},
			\end{aligned}
		\end{equation}	
		where $(k,p)\in\mathcal{R}$ represents the DD region far from $(\tilde{k},\tilde{p})$. Note that the approximation in (\ref{equ160}) holds because the DR is typically much greater than 1, owing to the $NM$-fold accumulation of the target-related component, whereas the signaling randomness and the noise do not experience such accumulation.
		Overall, the DR refers to the power ratio of the strongest and the weakest targets in the delay-Doppler profile that can be detected \cite{hakobyan2017novel}, which may be approximated as the ratio between the peak and the pedestal of DD profiles in (\ref{equ160}). However, its relationship with other sensing metrics including the MSE of sensing CSI, $\text{SNR}_\text{out}$ and ISLR, remains unclear and will be thoroughly investigated in Sec. \ref{sec3}.

	\begin{figure}[!t]
		\centering
		\includegraphics[width=0.75\linewidth]{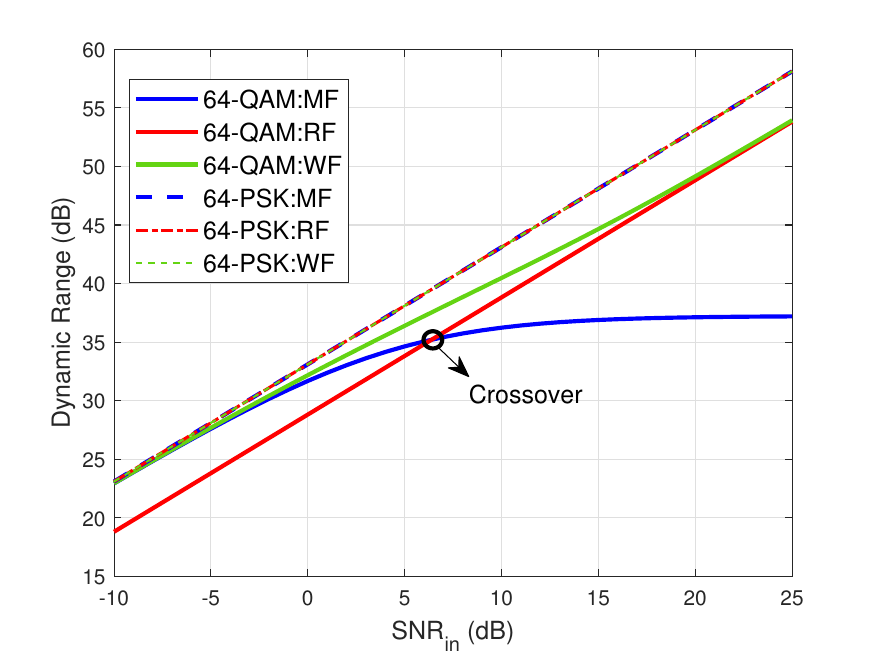}
		\caption{Dynamic range versus $\text{SNR}_\text{in}$ under different filtering schemes, given $N=64$ and $M=32$.}
		\label{figure10}
	\end{figure}

	Next, Fig. \ref{figure10} presents the DR versus $\text{SNR}_\text{in}$ under various filtering schemes and constellation types. In conjunction with (\ref{eq2}), several key observations can be summarized as follows.
	
	\begin{remark}
		If PSK is exploited, $\text{Var}\left(\chi\right)=0$ and $\frac{\mathbb{E}\{|g|^2\}}{\mathbb{E}^2\{\chi\}}=1$ for MF, RF and WF, accounting for the fact that these three filtering schemes exhibit the same DR.
		Therefore, an ideal DD profile requires adherence to the constant modulus constraint.
	\end{remark}

	\begin{remark}		 
		If non-constant modulus constellation, e.g., QAM, is exploited, MF and WF converge when $\text{SNR}_\text{in}$ is very low and diverges with increased $\text{SNR}_\text{in}$. To be more specific, three cases can be discussed:
		\begin{itemize}
			\item Medium-$\text{SNR}_\text{in}$ Case: This can represent a general case, since the effect of signaling randomness and filtering schemes on the sensing performance can be quantified by $\text{Var}\left(\chi\right)$, $\mathbb{E}^2\{\chi\}$ and $\mathbb{E}\{|g|^2\}$, in accordance with (\ref{eq2}).
			Interestingly, in this case RF and MF can have a crossover point when they achieve the same DR, i.e.,
			\begin{equation}
				\begin{aligned}
					\sigma^2 \mathbb{E}\left\{ {|x|^{-2}} \right\} =\sigma^2_{\alpha}\left(\mathbb{E}\left\{|x|^4\right\}-1\right)+\sigma^2,
				\end{aligned}
			\end{equation}
			or equivalently,
			\begin{equation}\label{e21}
				\begin{aligned}
					\text{SNR}_\text{in}=\frac{\mathbb{E}\left\{ {|x|^{-2}} \right\}-1}{\mathbb{E}\left\{|x|^4\right\}-1}.
				\end{aligned}
			\end{equation}
			For uniform 64-QAM, we can calculate (\ref{e21}) as $\text{SNR}_\text{in}=6.45$ dB. This is also verified in Fig. \ref{figure10}, where a cross point between MF and RF appears nearby $\text{SNR}_\text{in}=6.45$ dB.
			\item High-$\text{SNR}_\text{in}$ Case ($\sigma^2\rightarrow 0$): $\mathbb{E}^2\{\chi\}=1$ and $\sigma^2\mathbb{E}\{|g|^2\}=0$ hold for three filtering schemes. In contrast, the DD profile of MF would be deteriorated since $\text{Var}\left(\chi\right)>0$, while $\text{Var}\left(\chi\right)=0$ for RF and WF. Therefore, RF and WF converge and achieve better sensing performance than MF.
			\item Low-$\text{SNR}_\text{in}$ Case ($\sigma^2\rightarrow +\infty$): Evidently, in this case the sensing performance is mainly determined by $\sigma^2\mathbb{E}\{|g|^2\}$.
			First, MF and WF are equivalent. This can be explained by referring to $\text{SNR}_\text{in}^{-1}\gg \left\vert x \right\vert^2$, leading to  the fact that $g=\frac{x^*}{\left\vert x \right\vert^2+\text{SNR}_\text{in}^{-1}}\approx \text{SNR}_\text{in}x^*$. Therefore, the DR of WF can be approximated as 
			\begin{equation}
				\begin{aligned} 
					& \frac{NM\sigma^2_{\alpha}\mathbb{E}^2\{\chi\}}{\sigma^2\mathbb{E}\{|g|^2\}}  =\frac{NM\text{SNR}_\text{in}\mathbb{E}^2\left\{\frac{|x|^2}{|x|^2+\text{SNR}^{-1}_\text{in}}\right\}}{\mathbb{E}\left\{\frac{|x|^2}{(|x|^2+\text{SNR}^{-1}_\text{in})^2}\right\}} \\ & \qquad \approx NM\text{SNR}_\text{in} \mathbb{E}\left\{|x|^2\right\}=NM\text{SNR}_\text{in},
				\end{aligned}
			\end{equation}
			which equals to the DR of MF.
			Second, MF and WF behave better than RF, because RF significantly amplifies the noise power $\sigma^2\mathbb{E}\{|g|^2\}$ than MF and WF.
		\end{itemize}
	\end{remark}

\begin{figure}[!t]
		\centering
		\includegraphics[width=0.75\linewidth]{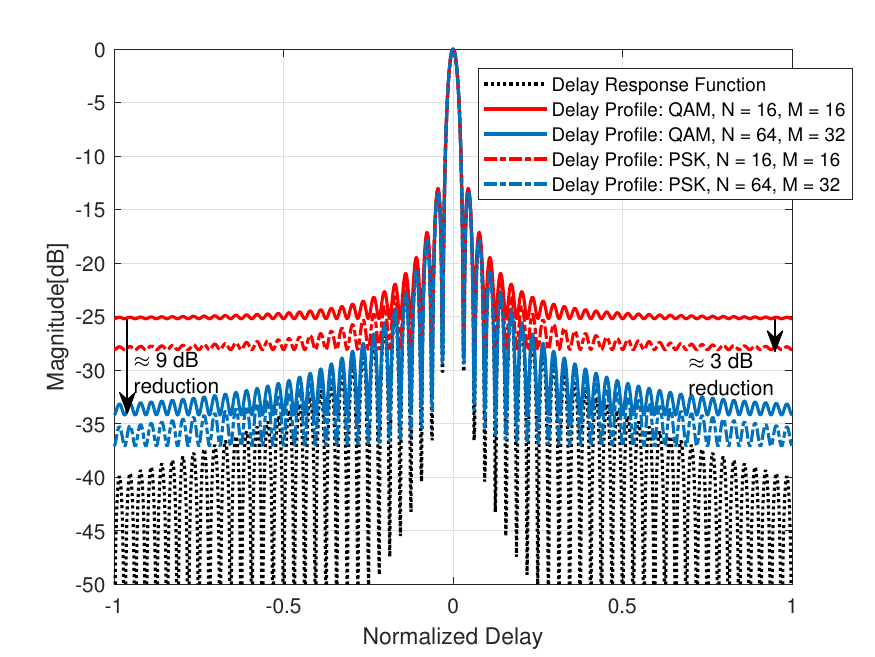}
		\caption{Normalized zero-Doppler slice of $\mathbb{E}\{|\bf{\Lambda}|^2\}$ ($\tilde{k}=0$, $\tilde{p}=0$ and $\text{SNR}_\text{in}=4$ dB) with different numbers of subcarriers and symbols.} 
		\label{figure1}
	\end{figure}

	\begin{remark}
		As shown in (\ref{eq2}), coherently integrating subcarriers and symbols enhances the sensing performance. For instance, increasing $N$ from 16 to 64 and $M$ from 16 to 32 improves the overall DD profile by approximately $10\log_{10}\frac{64\times32}{16\times16}\approx9$ dB, as demonstrated in Fig. \ref{figure1} which is exemplified by MF. In Fig. \ref{figure1}, the DD response function corresponds to the second term (of interest) in the last equation of (\ref{eq2}), and the range profile can be raised due to the additional corruption of $\sigma^2_{\alpha}{\text{Var}\left(\chi\right)}+\sigma^2\mathbb{E}\{|g|^2\}$ (i.e., the pedestal of DD profiles). Zero-delay slice is analogous and thus omitted due to the space limitation. Moreover, exploiting PSK can further obtain $\approx3$ dB performance gain relative to QAM, revealing the potential of introducing PCS in achieving a flexible trade-off between PSK and QAM. Notably, the results in Fig. \ref{figure1} align with the ``iceberg-in-the-sea'' structure described in \cite{liu2025uncovering,liu2025sensing}. In our findings, the ``iceberg'' represents the normalized response function with a sinc structure in the DD domain, while the ``sea level'' is contributed by both signaling randomness and output noise which may be amplified. Evidently, with a larger $NM$, the ``sea level'' can be suppressed to enable more sidelobes of ``iceberg'' to emerge, indicating that PCS has a weaker impact on scaling the sensing performance near the mainlobe region. In particular, as $NM\rightarrow+\infty$, the effects of signaling randomness and noise become negligible in the DD profile. However, increasing $N$ makes OFDM signaling more sensitive to Doppler shift and phase noise, while increasing $M$ does not necessarily ensure coherent S\&C processing in high-mobility environments. Therefore, on the premise of practical system parameters, PCS may effectively mitigate the effects of signaling randomness and noise amplification.
	\end{remark}

	\subsection{Communication Signal Model}
	
	The received OFDM communication signal at the 
	$n$th subcarrier and the $m$th OFDM symbol can be expressed as  $y^c_{n,m}=h^c_{n,m}x_{n,m}+z^c_{n,m}$, with its matrix-vector version over the entire OFDM frame as \cite{keskin2024fundamental}
	\begin{equation}
		\begin{aligned}
			\bf{Y}_c= \bf{H}_c\odot\bf{X}+\bf{Z}_c,
		\end{aligned}
	\end{equation}
	where $\bf{H}_c\in\mathbb{C}^{N\times M}$ and $\bf{Z}_c\in\mathbb{C}^{N\times M}$ represent the doubly selective fading channel matrix and the noise with zero mean and power $\sigma^2_C$, respectively.
	
	The mutual information is used to measure the AIR, which is expressed as \cite{du2024reshaping,keskin2024fundamental,du2019information}
	\begin{equation}\label{equ1}
		\begin{aligned}
			I\left(\bf{X};\bf{Y}_c|\bf{H}_c\right) = \sum\nolimits_{n,m} I\left(x_{n,m};y^c_{n,m}|h^c_{n,m}\right),
		\end{aligned}
	\end{equation}
	where
	\begin{equation}\label{equ1}
		\begin{aligned}
			I\left(x_{n,m};y^c_{n,m}|h^c_{n,m}\right) = H(y^c_{n,m}|h^c_{n,m})-H(y^c_{n,m}|h^c_{n,m},x_{n,m}).
		\end{aligned}
	\end{equation}
	
	Clearly, the conditional entropy $H(y^c_{n,m}|h^c_{n,m},x_{n,m})=\log \left(\pi e \sigma^2_C\right)$, corresponding to the entropy of the Gaussian noise. In contrast, $H(y^c_{n,m}|h^c_{n,m})$ has no closed-form due to the mixture Gaussian distribution of $y^c_{n,m}$ \cite{du2024reshaping}. We therefore approximately compute $H(y^c_{n,m}|h^c_{n,m})$ using Monte Carlo numerical integration:
	\begin{equation}\label{Eq15}
		\begin{aligned}
			H(y^c_{n,m} & |h^c_{n,m}) = -\mathbb{E} \left[ \log \sum\nolimits_{x} p(y^c_{n,m}|h^c_{n,m},x_{n,m}) p(x) \right] \\
			\approx & -\frac{1}{L}\sum\nolimits^{L}_{l=1} \log \sum\nolimits_{x} p\left(y^c_{n,m}(l)|h^c_{n,m},x\right) p(x),
		\end{aligned}
	\end{equation}
	where $L$ represents the number of Monte Carlo trials, $y^c_{n,m}(l)$ denotes the $l$th observation and its conditional probability density function $p\left(y^c_{n,m}(l)|h^c_{n,m},x\right)$ is with standard Gaussian forms in the $l$th trial, for each $x$ in the given constellation. By leveraging this approximation, the AIR can be accurately observed and exploited in the subsequent PCS optimization.
	
	\subsection{PCS Model for ISAC}\label{subsec4}
	
	By maximizing the communication metric (i.e., the mutual information) under the constraints of sensing metric (i.e., the DR), probability and transmit power, we hereby present the unified PCS approach compatible to MF and MMF schemes, formulated as follows:
	\begin{equation}\label{equ36}
		(\mathcal{P}1)
		\left\{
		\begin{aligned}
			\max_{\bf{p}_x}  & \ I\left(x;y_c|h_c\right)
			\\
			s.t. \ & C_1: \frac{1}{\text{DR}} \leq c_0, \
			C_2: \sum\nolimits_{x}p(x)|x|^2 = 1, \\
			& C_3: \sum\nolimits_{x} p(x)= 1, \
			C_4: 0\leq p(x) \leq 1 
		\end{aligned}
		\right.
	\end{equation}
	where $p(x)$ denotes the input distribution of $\tilde{Q}$ discrete symbols drawn from a given QAM constellation, with its discrete counterpart as $\bf{p}_x = [p_{x,1}, p_{x,2},...,p_{x,\tilde{Q}}]^T$. In addition, $c_0$ denotes a pre-determined constant that controls the DR.
	
	However, the objective function has no closed-form expression. Therefore, the model cannot be efficiently solved via existing numerical tools, e.g., CVX toolbox \cite{grant2014cvx}.
	To tackle this optimization problem, we first reformulate the objective in (\ref{equ36}) as \cite{yeung2008information}
	\begin{equation}\label{equ39}
		\begin{aligned}
			& \max_{\bf{p}_x}  \ I\left(x;y_c|h_c\right) = \max_{\bf{p}_x} \max_{\bf{q}_{x|y_c}} \ \mathcal{F}(\bf{p}_x,\bf{q}_{x|y_c}),
		\end{aligned}
	\end{equation}
	where $\mathcal{F}(\bf{p}_x,\bf{q}_{x|y_c})=\sum\nolimits_{x} \sum\nolimits_{y_c} p(x)p(y_c|x)\log\frac{q(x|y_c)}{p(x)}$, and $q(x|y_c)$ is the probability transition function from the input alphabet to the output alphabet with its discrete version as $q(x|y_c)$. 
	
	Next, one may hope to solve the optimization model through a modified Blahut-Arimoto (MBA) algorithm \cite{du2024reshaping} by referring to \cite{yeung2008information}, inspired by the original idea of alternating optimization between $\bf{p}_x$ and $\bf{q}_{x|y_c}$. 
	However, $C_1$ can be rewritten as $\frac{\sigma^2_{\alpha}\text{Var}\left(\chi\right)+\sigma^2\mathbb{E}\{|g|^2\}}{NM\sigma^2_{\alpha}\mathbb{E}^2\{\chi\}} \leq c_0$, which is equivalent to $\sigma^2_{\alpha}(1+c_0NM)\mathbb{E}^2\{\chi\}-\sigma^2_{\alpha}\mathbb{E}\{\chi^2 \}-\sigma^2\mathbb{E}\{|g|^2\} \geq 0$. This inequality does not define a convex set, thereby rendering the MBA algorithm proposed in \cite{du2024reshaping} inapplicable.

%

	\section{Sensing Performance Metrics}\label{sec3}
	The aforementioned DR can be adopted as a sensing performance indicator, while it is non-applicable for PCS optimization. This limitation compels us to seek for another sensing metric. In specific, given a DD profile generated by a filter, celebrated sensing metrics include the $\text{SNR}_\text{out}$ and ISLR. Besides, the sensing performance naturally depends on the estimation accuracy of sensing CSI. However, their analytical relationships remain un-characterized. To proceed, we first derive the MSE of sensing CSI, $\text{SNR}_\text{out}$, and ISLR, for MF, RF and WF, respectively. Then we explore the relationship among these three sensing criteria, and depict why the MSE is a more comprehensive criterion to evaluate the filtering result, in particular its superiority in constructing a convex model for PCS.
	
	\subsection{MSE of Sensing CSI}
	The MSE of sensing CSI can be defined as \cite{kay1993fundamentals}
	\begin{equation}\label{mse}
		\begin{aligned}
			\mathbb{E}\{\varepsilon^2\} = \mathbb{E} \left\{\left\Vert\hat{\bf{H}}-\bf{H}\right\Vert^2_F\right\},
		\end{aligned}
	\end{equation} 
	where the expectation is defined over the random variables $\bf{X}$, $\bf{H}$, and $\bf{Z}$. 
	
	Inserting (\ref{csi}) into (\ref{mse}) reformulates the sensing CSI into
		\begin{equation}\label{equ41}
			\begin{aligned}
				& \mathbb{E} \{\varepsilon^2\} = \mathbb{E} \left\{\left\Vert \bf{H}\odot\bf{X}\odot\bf{G}-\bf{H}+\bf{Z}\odot\bf{G}\right\Vert^2_F\right\} \\ & = \mathbb{E} \left\{\left\Vert \bf{H}\odot\left(\bf{X}\odot\bf{G}-\bf{1}_N\bf{1}^T_M\right)\right\Vert^2_F\right\} +\mathbb{E} \left\{\left\Vert \bf{Z}\odot\bf{G}\right\Vert^2_F\right\},
			\end{aligned}
		\end{equation}
	where the second equality holds due to the statistical independence between $\bf{H}$ and $\bf{Z}$.
	
	Due to the fact that the statistical characteristics of random variables in (\ref{equ41}) are invariant across subcarriers and symbols, the MSE can be further simplified as
	\begin{equation}\label{equ42}
		\begin{aligned}
			\mathbb{E} \{\varepsilon^2\} = & \mathbb{E} \left\{\sum_{n,m}|h_{n,m}|^2\left( \chi_{n,m}-1\right)^2\right\} + \mathbb{E} \left\{\sum_{n,m}|z_{n,m}g_{n,m}|^2\right\}  \\ = & \sigma^2_{\alpha} \mathbb{E} \left\{\sum_{n,m}\left( \chi_{n,m}-1\right)^2\right\} + \sigma^2\mathbb{E} \left\{\sum_{n,m}|g_{n,m}|^2\right\}.
		\end{aligned}
	\end{equation}
	
	Substituting (\ref{equ26}) into (\ref{equ42}), we can obtain the MSE of MF, RF and WF, respectively, which are expressed in Table. \ref{tab}.

	\renewcommand{\arraystretch}{1.1} 
	\begin{table*}[h]
		\centering
		\caption{The MSE of sensing CSI, $\text{SNR}_\text{out}$ and ISLR, among MF, RF, and WF.}
		\Large 
		\resizebox{1.6\columnwidth}{!}{
			\begin{tabular}{c|c|c|c} 
				\hline
				\hline
				& MF & RF & WF \\
				\hline 
				$\mathbb{E}(\varepsilon^2)$ & $NM\left( \sigma^2_{\alpha} \left(\mathbb{E}\left\{|x|^4\right\}-1\right) + \sigma^2 \right)$ & $NM\sigma^2\mathbb{E}\left\{ \frac{1}{|x|^2} \right\}$ &  $NM\sigma^2 \mathbb{E}\left\{ \frac{1}{|x|^2+\text{SNR}_\text{in}^{-1}} \right\}$ \\
				\hline
				$\text{SNR}_\text{out}$ & $\text{SNR}_\text{in} \cdot \left(\mathbb{E}\left\{ \left\vert x \right\vert^4 \right\}+NM-1\right)$  &  $\text{SNR}_\text{in} \cdot \frac{NM}{\mathbb{E}\left\{ \frac{1}{|x|^2}\right\}}$ & $\text{SNR}_\text{in} \cdot \frac{\mathbb{E}\left\{ \frac{\left\vert x \right\vert^4}{\left(\left\vert x \right\vert^2+\text{SNR}_\text{in}^{-1}\right)^2} \right\} + (NM-1)\mathbb{E}^2\left\{ \frac{\left\vert x \right\vert^2}{\left\vert x \right\vert^2+\text{SNR}_\text{in}^{-1}} \right\}}{ \mathbb{E}\left\{\frac{\left\vert x \right\vert^2}{\left(\left\vert x \right\vert^2+\text{SNR}_\text{in}^{-1}\right)^2}\right\}}$ \\
				\hline 
				ISLR & $\frac{\left(NM-1\right)\left(\mathbb{E}\left\{ \left\vert x \right\vert^4 \right\}-1\right)}{NM+\left(\mathbb{E}\left\{ \left\vert x \right\vert^4 \right\}-1\right)}$ & $0$ & $\frac{\left(NM-1\right)\left(\mathbb{E}\left\{ \frac{|x|^4}{\left(|x|^2+\text{SNR}_\text{in}^{-1}\right)^2} \right\} - \mathbb{E}^2\left\{ \frac{|x|^2}{|x|^2+\text{SNR}_\text{in}^{-1}} \right\}\right)} {\mathbb{E}\left\{ \frac{|x|^4}{\left(|x|^2+\text{SNR}_\text{in}^{-1}\right)^2} \right\} + (NM-1)\mathbb{E}^2\left\{ \frac{|x|^2}{|x|^2+\text{SNR}_\text{in}^{-1}} \right\} }$ \\
				\hline
				\hline
			\end{tabular}
		}
		\label{tab}
	\end{table*}	
	
	\subsection{Output SNR}
	
	MF is optimal in terms of maximizing $\text{SNR}_\text{out}$, as formally proven using the Cauchy–Schwarz inequality \cite{kay1993fundamentals}. In contrast, MMF schemes inherently experience a loss in $\text{SNR}_\text{out}$. For ISAC systems employing OFDM random signals, $\text{SNR}_\text{out}$ is defined as
	\begin{equation}\label{snrout}
		\begin{aligned}
			\text{SNR}_\text{out} 
			= & \frac{\max\limits_{<p,k>} \ \mathbb{E}\left\{ \sigma^2_{\alpha}\left\vert r(k,p) \right\vert^2 \right\}}{\frac{1}{NM} \mathbb{E}\left\{\sigma^2 \sum_{n,m}|g_{n,m}|^2 \right\} }.
		\end{aligned}
	\end{equation}
	
	Recalling (\ref{equ27}) and (\ref{eq2}), it is evident that
	\begin{equation}\label{inequation}
		\begin{aligned}
			\mathbb{E}\left\{ \left\vert r(k,p) \right\vert^2 \right\} = & \text{Var}\left(\chi\right) + NM\mathbb{E}^2\{\chi\}\text{sinc}^2\left(k\right)\text{sinc}^2\left(p\right)
			\\ \leq & \text{Var}\left(\chi\right) + NM\mathbb{E}^2\{\chi\} = \mathbb{E}\left\{ r^2(0,0) \right\},
		\end{aligned}
	\end{equation}
	demonstrating that the numerator in (\ref{snrout}) is obtained when $k=0$ and $p=0$, where
	\begin{equation}\label{r00}
		\begin{aligned}
			r(0,0)=\frac{1}{\sqrt{NM}}\sum\nolimits_{n,m} \chi_{n,m}.
		\end{aligned}
	\end{equation}
	
	Accordingly, $\text{SNR}_\text{out}$ can be simplified as follows:
	\begin{equation}\label{snr}
		\begin{aligned}
			\text{SNR}_\text{out}  = & \frac{\sigma^2_{\alpha} \mathbb{E}\left\{r^2(0,0)  \right\}}{\frac{1}{NM}\sigma^2 \sum_{n,m}\mathbb{E}\left\{|g_{n,m} |^2 \right\} }
			\\ = & \text{SNR}_\text{in} \cdot \frac{\mathbb{E}\left\{ \left( \sum_{n,m} \chi_{n,m} \right)^2 \right\}}{\sum_{n,m} \mathbb{E}\left\{|g_{n,m} |^2 \right\} }.
		\end{aligned}
	\end{equation}
	Evidently, $\text{SNR}_\text{out}$ measures the sensing performance in terms of the mainlobe of $r(k,p)$ and the output noise power, regardless of sidelobes of $r(k,p)$.
		
	Plugging (\ref{equ26}) into (\ref{snr}) yields $\text{SNR}_\text{out}$ of MF, RF and WF, which are summarized in Table. \ref{tab}.

	\subsection{ISLR}
	ISLR is a popular sensing metric used to quantify the total energy of the sidelobes relative to the mainlobe of an $r(k,p)$, without involving the effect of noise. Therefore, ISLR can be formulated as 
	\begin{equation}\label{eq31}
		\begin{aligned}
			\text{ISLR} = \frac{ \mathbb{E} \left\{\sum_{k,p} |r(k,p)|^2\right\} -  \mathbb{E} \left\{r^2(0,0)\right\}} {\mathbb{E} \left\{r^2(0,0)\right\}}.
		\end{aligned}
	\end{equation}
	
	Next, we derive a variant of ISLR.
%

	\begin{corollary}
	The ISLR can also be reformulated as:
	\begin{equation}\label{eq33}
		\begin{aligned}
			\text{ISLR} = \frac{ \mathbb{E} \left\{\sum_{n,m} \left(\chi_{n,m}-r(0,0)/\sqrt{NM}\right)^2 \right\} }{\mathbb{E} \left\{r^2(0,0)\right\}}.
		\end{aligned}
	\end{equation}
	\end{corollary}
	
	\textbf{\textit{Proof:}}	
	See Appendix \ref{appendix2}.
	\hfill $\blacksquare$
	\vspace{2mm}

	Inserting (\ref{equ26}), (\ref{dft}) and (\ref{eqxxx}) into (\ref{eq31}), the ISLR of MF, RF and WF can thus be derived as in Table. \ref{tab}. It is worth mentioning that the ISLR of RF is zero, since the element-wise division yields a flat TF response function, corresponding to the delta-shaped DD response function with zero sidelobes.

	\subsection{Discussions}
	\subsubsection{Relationship among MSE, $\text{SNR}_\text{out}$ and ISLR}
	\begin{theorem}
		The relationship among the MSE of sensing CSI, the ISLR and the output SNR can be established as
		\begin{equation}\label{theorem}
			\begin{aligned}
				& {\sigma^2_{\alpha}}\left(\underbrace{\text{ISLR}+ \frac{NM \mathbb{E} \left\{\left(1-r(0,0)/\sqrt{NM}\right)^2\right\}}{\mathbb{E} \left\{r^2(0,0)\right\}}}_{r(k,p) \ \text{effect}} + \underbrace{\frac{1}{\text{SNR}_\text{out}}}_{\text{noise effect}}\right)  \\ & = \frac{\mathbb{E}\{\varepsilon^2\}}{\mathbb{E} \left\{r^2(0,0)\right\}}.
		\end{aligned}
		\end{equation}
	\end{theorem}
	
	\textbf{\textit{Proof:}}
	See Appendix \ref{appendix3}.
	\hfill $\blacksquare$
	\vspace{2mm}
	
	\begin{remark}
		Overall, (\ref{theorem}) reveals that $\mathbb{E}\{\varepsilon^2\}$ is contributed by three factors: the mainlobe loss of $r(k,p)$, the accumulation of sidelobes of $r(k,p)$, and the potential amplification of noise power. Therefore, $\mathbb{E}\{\varepsilon^2\}$ may serve as a more suitable sensing metric over ISLR and $\text{SNR}_\text{in}$. However, the mathematical relationship between $\mathbb{E}\{\varepsilon^2\}$ and the DD profile (\ref{eq2}) has not yet been exactly established. To address this, we present another theorem below.
	\end{remark}

	\subsubsection{Decomposition of the MSE in the DD domain}
	
	\begin{theorem}
		While the MSE in (\ref{equ41}) and (\ref{equ42}) is defined in the TF domain, it can also be interpreted as the accumulation of errors in the DD domain, which is expressed as
		\begin{equation}\label{mse2}
			\begin{aligned}
				\mathbb{E}\{\varepsilon^2\} = \text{Var}\left(\bf{F}^H_N{\left(\hat{\bf{H}}-\bf{H}\right)}\bf{F}_M\right),
			\end{aligned}
		\end{equation}
		where $\text{Var}(\bf{A}) = \mathbb{E}\{\left\Vert \bf{A}-\mathbb{E}\left\{\bf{A}\right\} \right\Vert^2_F\}$ represents the sum of entries in the matrix $\mathbb{E}\{\left|\bf{A}\right|^2\}-\left|\mathbb{E}\left\{\bf{A}\right\}\right|^2$.
	\end{theorem}
	
	\textbf{\textit{Proof:}} 
	See Appendix \ref{appendix4}.
	\hfill $\blacksquare$
	\vspace{2mm}

		Next, we examine how the MSE relates to the DD profiles characterized in (\ref{eq2}). To proceed, we reformulate (\ref{mse2}) as
		\begin{equation}\label{mse6}
			\begin{aligned}
				\mathbb{E}\{\varepsilon^2\} = &  \text{Var}\left(\bf{F}^H_N\left(\hat{\bf{H}}-\bf{H}\right)\bf{F}_M\right) 
				\\ = & \sum\nolimits_{k,p} \mathbb{E}\{|\tilde{\Lambda}_{k,p}|^2\} - \sum\nolimits_{k,p} |\mathbb{E}\{\tilde{\Lambda}_{k,p}\}|^2
				\\ = & \sum\nolimits_{k,p} \mathbb{E}\{|\tilde{\Lambda}_{k,p}|^2\},
			\end{aligned}
		\end{equation}
		where $\tilde{\Lambda}_{k,p}$ is the $(k,p)$th entry of $\tilde{\bf{\Lambda}} = \bf{F}^H_N\left(\hat{\bf{H}}-\bf{H}\right)\bf{F}_M$, and $\mathbb{E}\{\tilde{\Lambda}_{k,p}\}=0$ is exploited in (\ref{mse6}) for simplicity.
		Similar to derivations in (\ref{eq2}), one may obtain
%
		\begin{equation}\label{equ43}
			\begin{aligned}
				\mathbb{E} \{\varepsilon^2\} 
				= &  NM\sigma^2_{\alpha}\sum_{k,p}\left[\left(\mathbb{E}\{\chi\}-1\right)\text{sinc}\left(k-\tilde{k}\right)\text{sinc}\left(p-\tilde{p}\right)\right]^2 \\ 
				& + \sum\nolimits_{k,p}\sigma^2_{\alpha}\text{Var}\left(\chi-1\right) +\sum\nolimits_{k,p}\sigma^2\mathbb{E}\{|g|^2\}
				\\ = & NM\left( \sigma^2_{\alpha} \left(\mathbb{E}\{\chi^2 \} -2 \mathbb{E}\{\chi\}+1\right) +\sigma^2\mathbb{E}\{|g|^2\} \right)
				\\ = & (\ref{equ42})
			\end{aligned}
		\end{equation}
		where the second equal sign holds relying on $\text{Var}(\chi-1)=\text{Var}(\chi)$, and $\sum\nolimits_{k,p}\text{sinc}^2 (k-\tilde{k})\text{sinc}^2 (p-\tilde{p}) = 1$
		which holds both for on-grid and off-grid cases\footnote{This can be explained by referring to the squared magnitude sum property of the Dirichlet kernel, where the total energy in the 2D-DFT domain remains constant, even when the signal is off-grid. This follows directly from the Parseval identity applied to the DFT basis.}.
		The result in (\ref{equ43}) naturally implies the MSE of sensing CSI in the DD domain equals to that in the TF domain.
		
\begin{figure}[!t]
				\centering
				\includegraphics[width=0.8\linewidth]{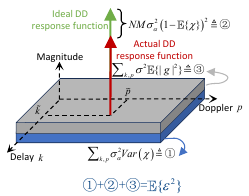}
				\caption{Illustration of relationship between $\mathbb{E} \{\varepsilon^2\}$ and $\mathbb{E}\{|\bf{\Lambda}|^2\}$. For clarity, the on-grid target is employed in this figure. In contrast, the green and red arrows should be replaced by sinc-shaped counterparts for the off-grid target.}
				\label{figure2}
			\end{figure}

		\begin{remark}
			We now specialize the relationship between $\mathbb{E} \{\varepsilon^2\}$ and (\ref{eq2}), as illustrated in Fig. \ref{figure2}.
			First, recalling (\ref{eq2}), $\text{Var}\left(\chi\right)$ and $\sigma^2\mathbb{E}\{|g|^2\}$ constitute the pedestal of DD profiles induced by the signaling randomness and the amplified noise, which are constant with respect to indices $k$ and $p$. Second, $\mathbb{E}\{\chi\}\text{sinc}\left(k-\tilde{k}\right)\text{sinc}\left(p-\tilde{p}\right)$ represents the DD response function (e.g., red arrow in Fig. \ref{figure2}), which may be smaller than the ideal  $\text{sinc}\left(k-\tilde{k}\right)\text{sinc}\left(p-\tilde{p}\right)$ (e.g., green arrow in Fig. \ref{figure2}) due to $\mathbb{E}\{\chi\}\leq 1$ for MF, RF, and WF, thereby leading to the squared error of $(1-\mathbb{E}\{\chi\})^2\sum_{k,p}\text{sinc}^2 \left(k-\tilde{k}\right)\text{sinc}^2 \left(p-\tilde{p}\right) = (1-\mathbb{E}\{\chi\})^2$.
		\end{remark}
			
			\subsubsection{Relating normalized MSE (NMSE) to DR} 
\begin{figure}[!t]
		\centering  
		\subfigure[MSE]{
			\includegraphics[width=2.39in]{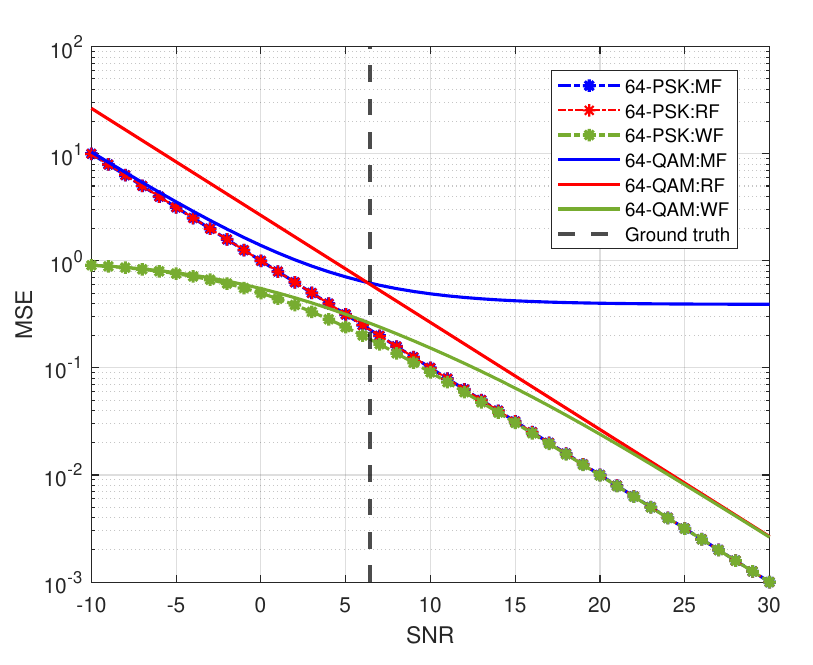}}
		\subfigure[NMSE]{
			\includegraphics[width=2.52in]{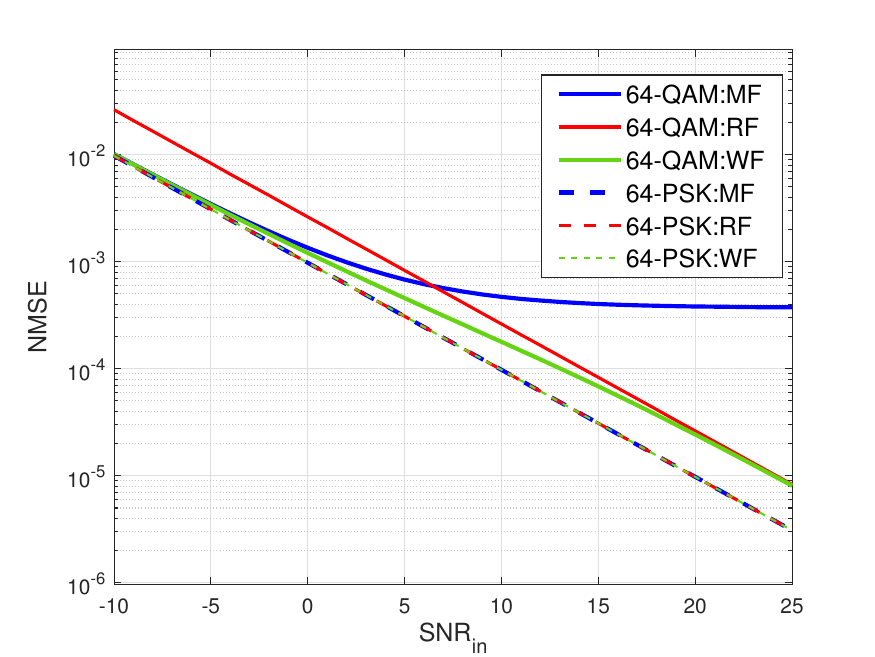}}
		\caption{MSE and NMSE versus $\text{SNR}_\text{in}$ under different filtering schemes.}\label{figure19}
	\end{figure}

The MSE is not suitable for fair performance comparison across different filters due to its scale-variant characteristic. For example, under PSK modulations, the MSE of WF remains smaller than those of MF and RF in the low $\text{SNR}_\text{in}$ regime (See Fig. \ref{figure19}(a)), even though all three filtering schemes theoretically achieve the same sensing performance in terms of the DR as depicted in Fig. \ref{figure10}. In contrast, the NMSE values of MF, RF, and WF are identical in this case (see Fig. \ref{figure19}(b)), providing a scale-invariant metric that enables consistent and meaningful performance comparisons across different filtering schemes. To that end, the NMSE can be defined as $\frac{\mathbb{E} \{\varepsilon^2\}}{\sigma^2_{\alpha}\mathbb{E}^2 \{\chi\}}$, which is reformulated as
			\begin{equation}\label{nmse}
				\begin{aligned}
					\frac{\mathbb{E} \{\varepsilon^2\}}{\sigma^2_{\alpha}\mathbb{E}^2 \{\chi\}} = & \frac{NM\left[ \sigma^2_{\alpha}\text{Var}(\chi) +\sigma^2\mathbb{E}\{|g|^2\} \right]}{\sigma^2_{\alpha}\mathbb{E}^2 \{\chi\}} + \frac{\left(\mathbb{E}\{\chi\}-1\right)^2}{\mathbb{E}^2 \{\chi\}}
					\\
					= & \frac{N^2M^2}{\text{DR}} + \frac{\left(\mathbb{E}\{\chi\}-1\right)^2}{\mathbb{E}^2 \{\chi\}}.
				\end{aligned}
			\end{equation}
			If we employ the DR as the sensing metric, then optimizing the sensing performance can be formulated as
			\begin{equation}\label{dr}
				\begin{aligned}
					\min_{p(x)}  \ \frac{1}{\text{DR}}, \ \
					s.t. \ \mathbb{E}\{\chi\}\leq 1,
				\end{aligned}
			\end{equation} 
			where the constraint holds for MF, RF and WF and can be readily verified according to the expectation of (\ref{equ26}).
			 
			Combining (\ref{nmse}) and (\ref{dr}), we observe that the constrained optimization problem (\ref{dr}) may be recast as an unconstrained counterpart as
			\begin{equation}
				\begin{aligned}
					\min_{p(x)}  & \ \frac{1}{\text{DR}} + \rho\phi(\chi) \ \Leftrightarrow \min_{p(x)} \ \text{NMSE},
				\end{aligned}
			\end{equation} 
			where the penalty function and the penalty parameter are $\phi(\chi) = \left(\frac{\mathbb{E}\{\chi-1\}}{\mathbb{E}\{\chi\}}\right)^2$ and $\rho=\frac{1}{N^2M^2}$, respectively. In other words, the NMSE is a penalty function version of the DR. For clarity, the NMSE versus $\text{SNR}_\text{in}$ under various filtering schemes and constellation types, is also given in Fig. \ref{figure19}(b), which coincides with Fig. \ref{figure10}.


	

%

	\section{Unified PCS Approach for ISAC}\label{sec4}
	Based on the above analysis, the NMSE emerges as a preferable sensing metric for PCS optimization. However, incorporating the NMSE into constraint $C_1$ still results in a non-convex feasible set, similar to the case with the DR elaborated in Sec. \ref{subsec4}. Consequently, it is natural to adopt the MSE\footnote{Although the MSE value itself is not suitable for directly comparing the performance of different filters, it does depend on the constellation input probability for each individual filter. This justifies the use of the MSE in $\mathcal{P}2$, as our objective is to optimize the PCS scheme for each filter based on its own MSE behavior.} as the sensing metric, leading to the PCS model as:
	\begin{equation}
		(\mathcal{P}2)
		\left\{
		\begin{aligned}
			\max_{\bf{p}_x} & \max_{\bf{q}_{x|y_c}} \ \mathcal{F}(\bf{p}_x,\bf{q}_{x|y_c})
			\\
			s.t. \ & C_1: \mathbb{E} \{\varepsilon^2\} \leq c_0, \\
			& C_2, \
			C_3, \
			C_4.
		\end{aligned}
		\right.
	\end{equation}
	The model $\mathcal{P}2$ is therewith a convex programming problem since it maximizes a jointly concave function with respect to $\bf{p}_x$ and $\bf{q}_{x|y_c}$, where all  constraints are linear. Then, we may exploit the MBA algorithm to solve $\mathcal{P}2$, with major steps summarized as follows.
	\begin{itemize}
		\item[1)] Given the $\ell$th iterative $\bf{p}^{(\ell)}_x$, maximizing the mutual information yields the updated $\bf{q}^{(\ell)}_{x|y_c}$ \cite{yeung2008information}:
		\begin{align}\label{iteration_formats}
			q^{(\ell)}(x|y_c) = & \frac{p^{(\ell)}(x)p(y_c|x)}{\sum_{x'}p^{(\ell)}(x')p(y_c|x')}.
		\end{align}
		\item[2)] Given the $\ell$th iterative $\bf{q}^{(\ell)}_{x|y_c}$, maximizing the mutual information yields the updated $\bf{p}^{(\ell+1)}_x$:
		\begin{align}
			p^{(\ell+1)}(x) = \frac{e^{\sum_{y_c} p(y_c|x)\log q^{(\ell)}(x|y_c) -\lambda_1 f(x) -\lambda_2|x|^2 } } {\sum_{x}e^{\sum_{y_c}p(y_c|x)\log q^{(\ell)}(x|y_c) -\lambda_1 f(x) -\lambda_2|x|^2 }},
		\end{align}
		where recalling (\ref{equ42}) yields
		\begin{equation}
			\begin{aligned}
				\frac{1}{NM\sigma^2} f(x) & \triangleq \left\{\text{SNR}_\text{in}\mathbb{E}\left\{ (\chi-1)^2\right\}+\mathbb{E}\left\{|g|^2\right\}\right\}
				\\ & = 
				\left\{
				\begin{array}{ll}
					\text{SNR}_\text{in}\left(|x|^4-1\right) +1, & \text{MF}, \\
					\frac{1}{|x|^2} , & \text{RF}, \\
					\frac{1}{\left\vert x \right\vert^2+\text{SNR}_\text{in}^{-1}}, & \text{WF}
				\end{array}
				\right.
			\end{aligned}
		\end{equation}
		Note that Lagrange multipliers $\lambda_1$ and $\lambda_2$ may be numerically solved with mature numerical algorithms, e.g., Newton's method.
		See \cite{du2024reshaping} for more technical details.
		\item[3)] Repeat until convergence, i.e., $\left\Vert\bf{p}^{(\ell+1)}_x-\bf{p}^{(\ell)}_x\right\Vert^2\leq \varepsilon_0$, where $\varepsilon_0$ denotes a tolerance, e.g., $\varepsilon_0=10^{-5}$.
	\end{itemize}

	\begin{remark}
		The determination of $c_0$: The effective range of $c_0$ in $\mathcal{P}2$ can be determined by respectively calculating $\mathbb{E}\{\varepsilon^2\}$ in (\ref{equ42}) for uniform PSK and QAM constellations, given $\sigma^2_{\alpha}$ and $\sigma^2$, under different filtering schemes. For a given QAM codebook, as $c_0$ is varied from the uniform PSK value (sensing-best) to the uniform QAM value (sensing-worst), it is expected that the communication performance will scale from the PSK’s AIR (communication-worst) to that of QAM (communication-best) \cite{du2024reshaping}. 
		Therefore, the desired S\&C trade-off can be controlled by selecting $c_0$ in this range.
	\end{remark}

	\begin{figure*}[!t]
		\centering 
		\subfigure[$\text{SNR}_\text{in}$=-10 dB]{
			\includegraphics[width=2.20in]{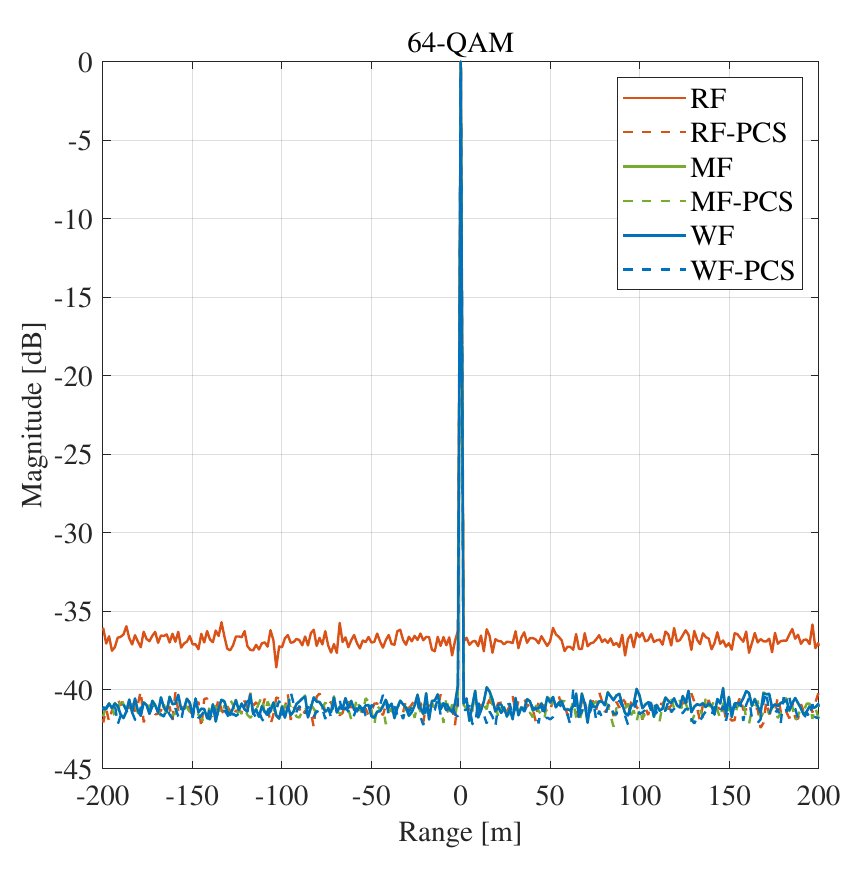}\label{figure3a}}
		\subfigure[$\text{SNR}_\text{in}$=4 dB]{
			\includegraphics[width=2.20in]{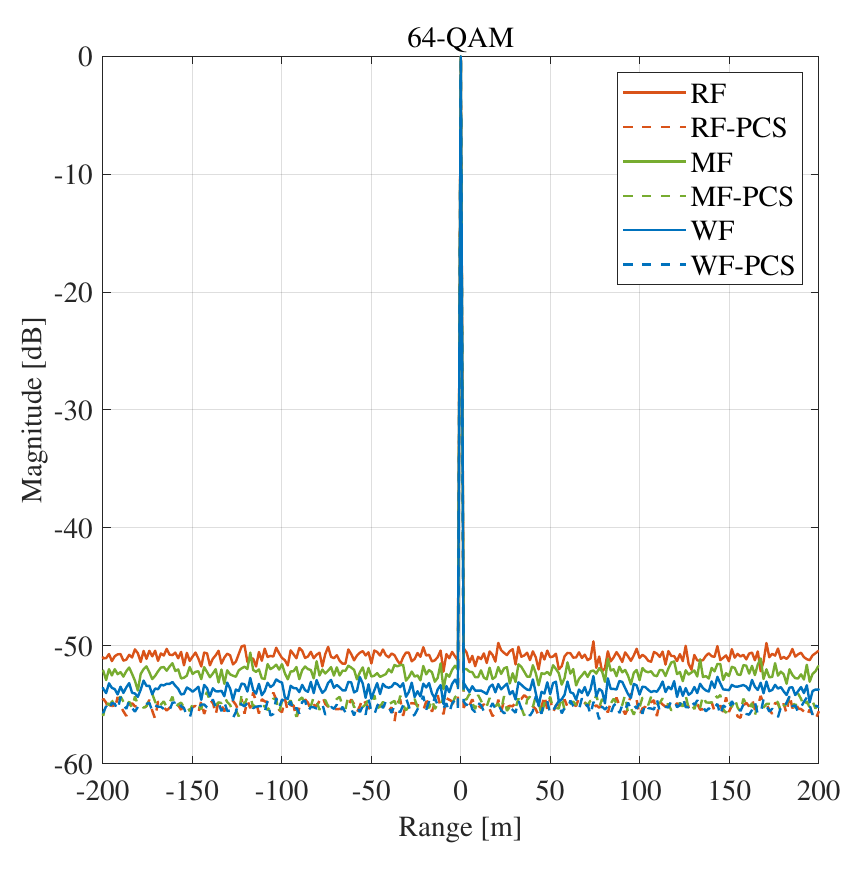}\label{figure3b}}
		\subfigure[$\text{SNR}_\text{in}$=20 dB]{
			\includegraphics[width=2.28in]{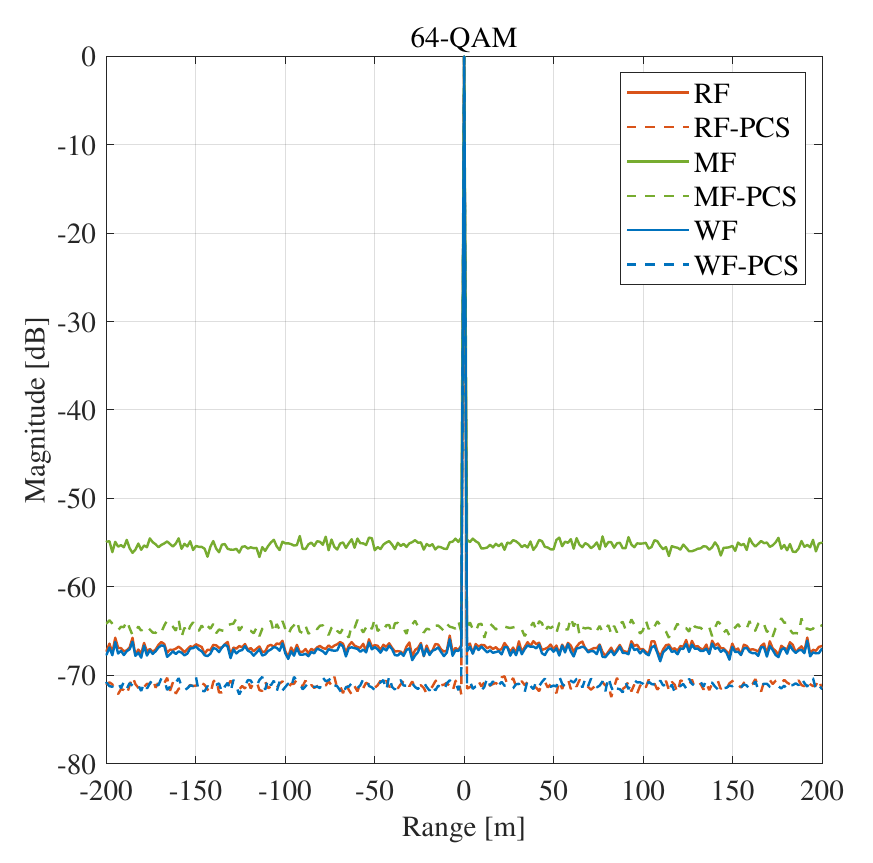}\label{figure3c}}
		\caption{Range profiles of MF/RF/WF with/without PCS optimizations under different input SNR values: simulation results.}\label{figure3}
	\end{figure*}

	\begin{figure*}[!t]
		\centering 
		\subfigure[$\text{SNR}_\text{in}$=-10 dB]{
			\includegraphics[width=2.24in]{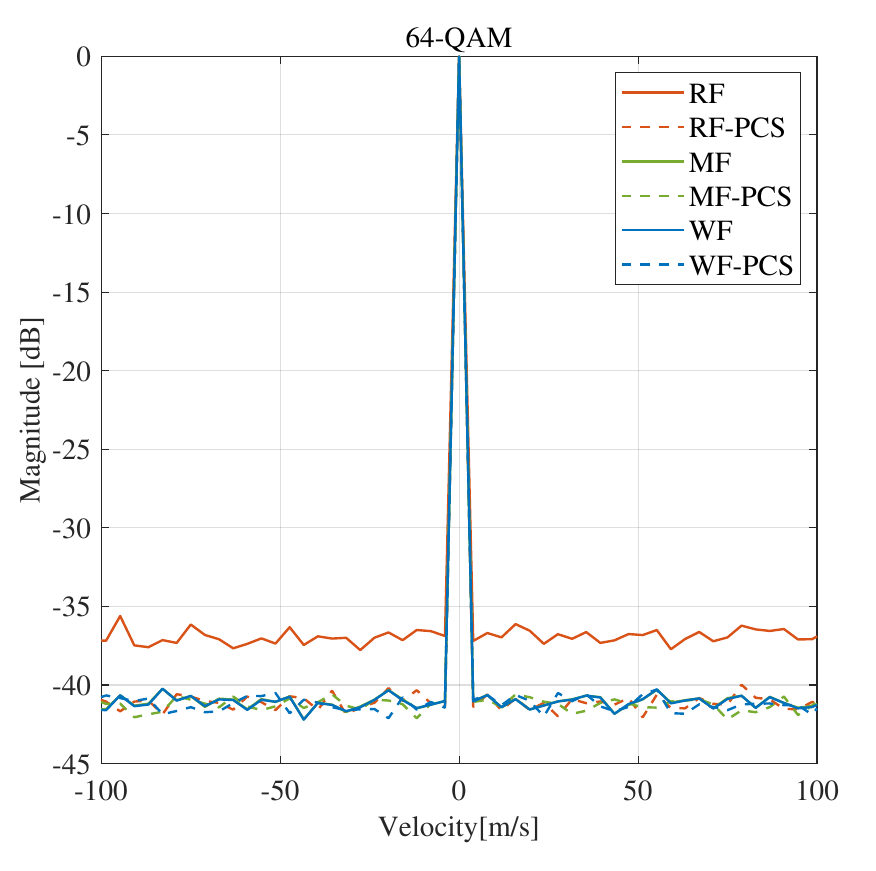}}
		\subfigure[$\text{SNR}_\text{in}$=4 dB]{
			\includegraphics[width=2.2in]{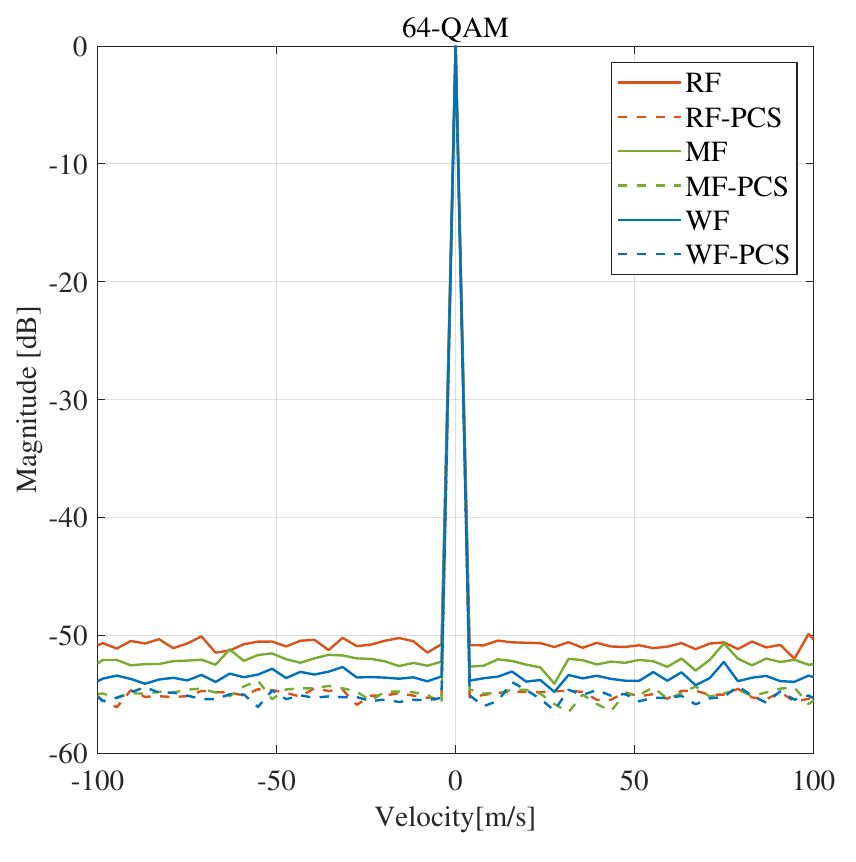}}
		\subfigure[$\text{SNR}_\text{in}$=20 dB]{
			\includegraphics[width=2.3in]{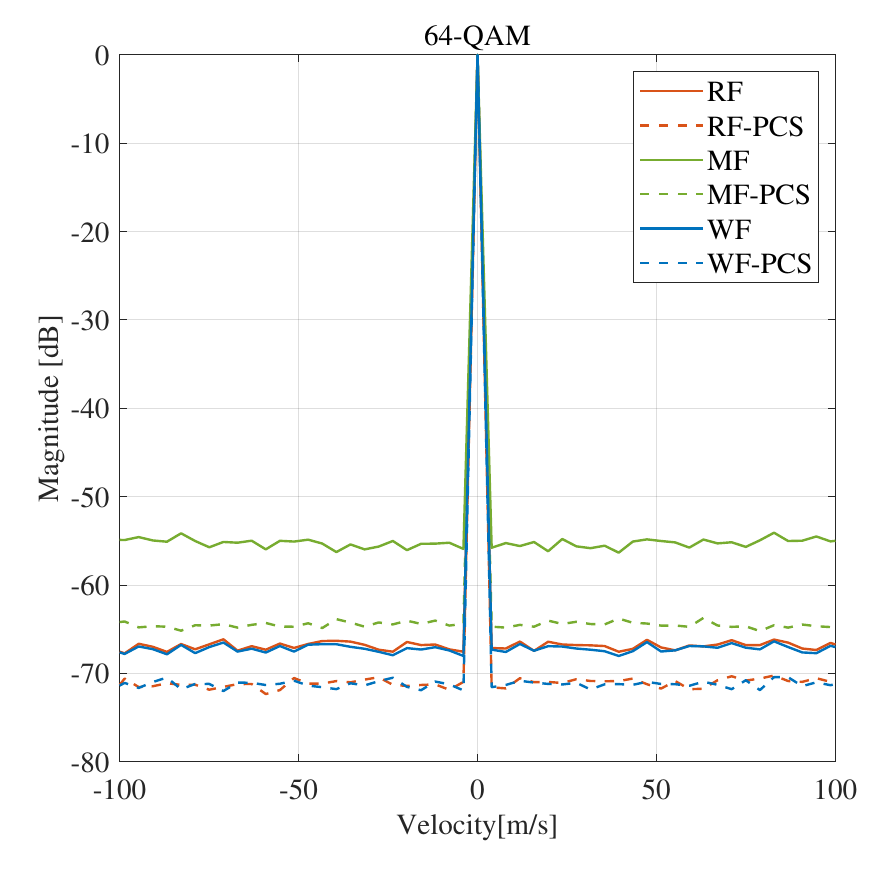}}
		\caption{Velocity profiles of MF/RF/WF with/without PCS optimizations under different input SNR values: simulation results.}\label{figure4}
	\end{figure*}

	\section{Performance Evaluation}\label{sec5}
	We aim to demonstrate the efficiency of PCS optimization in enhancing sensing performance compared to conventional QAM-constellation-modulated OFDM signaling, as well as its superiority in achieving a flexible S\&C trade-off. We commence with simulations and subsequently validate the benefits of our proposed approach through field measurements.
	
	\subsection{Sensing Performance Enhancement With PCS}
	Fig. \ref{figure3} presents the range profiles of MF, RF, and WF under $\text{SNR}_\text{in}$ of -10 dB, 4 dB and 20 dB, respectively, both with and without PCS optimization applied to a 64-QAM constellation. Unless otherwise specified, the PCS results are obtained using the specified $c_0$ that yields the best sensing performance. Without loss of generality, a static point target is assumed at zero range. Then the key findings are summarized as follows.
	
	\subsubsection{Low $\text{SNR}_\text{in}$ Case}
	As evidenced in Remark 2, the filtering results in this case are mainly determined by the output noise power $\sigma^2\mathbb{E}\{|g|^2\}$.
	When $\text{SNR}_\text{in}=-10$ dB in Fig. \ref{figure3a}, MF and WF (without PCS) achieve similar range profiles, significantly outperforming RF when exploiting uniform QAM modulations. This is because the output noise is considerably amplified due to the element-wise division imposed on non-constant modulus symbols with RF. In contrast, when PCS is exploited, RF optimizes the constellation toward a constant-modulus modulation, thereby leading to the best sensing performance as the noise power is not amplified either. 
		
	\subsubsection{Medium $\text{SNR}_\text{in}$ Case}	
	As depicted in Fig. \ref{figure3b} where $\text{SNR}_\text{in}=4$ dB, the sensing performance is simultaneously affected by the signaling randomness and the amplified noise power. Fortunately, exploiting PCS can almost improve 2 dB - 4 dB performance gain. This benefits from the fact that QAM constellations are optimized with PCS toward a constant-modulus modulation, as elaborated above, thereby alleviating the signaling randomness and the noise amplification.
		
	\subsubsection{High $\text{SNR}_\text{in}$ Case}
\begin{figure}[!t]
		\centering  
		\subfigure[MF]{
			\includegraphics[width=3.10in]{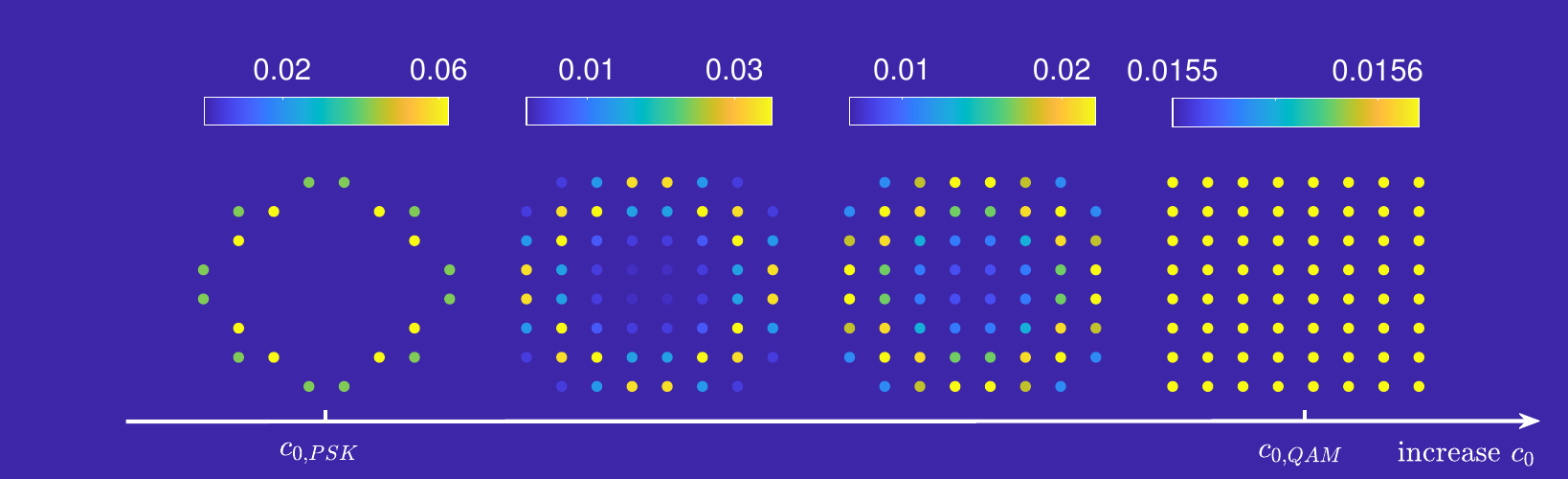}}
		\subfigure[RF]{
			\includegraphics[width=3.10in]{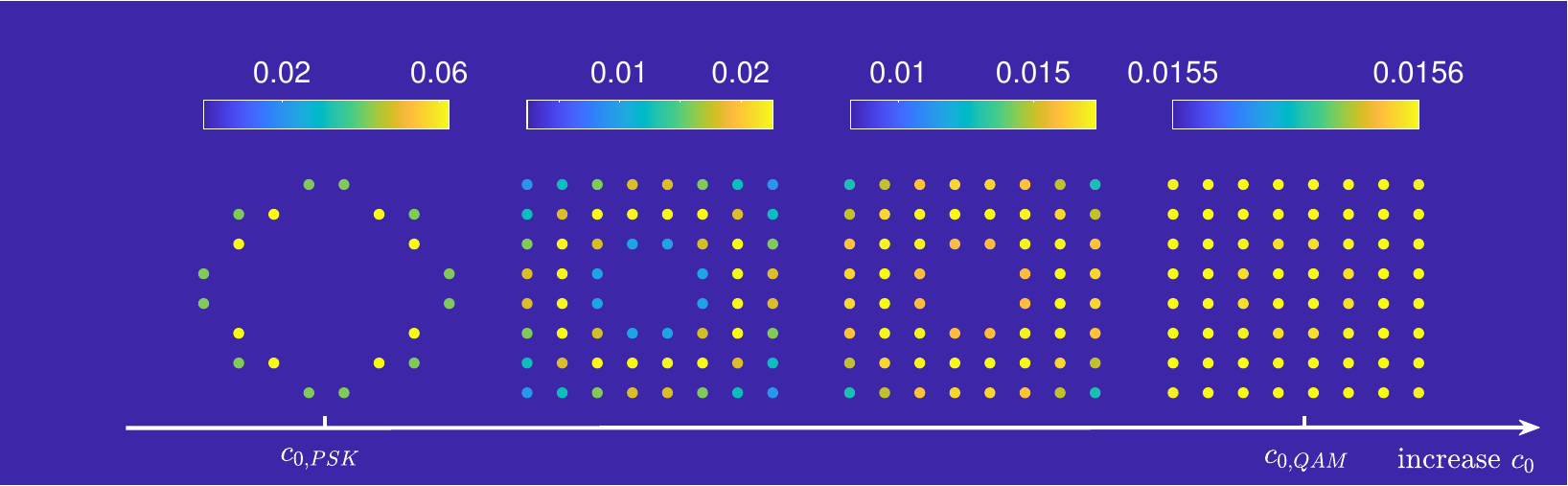}}
		\subfigure[WF]{
			\includegraphics[width=3.10in]{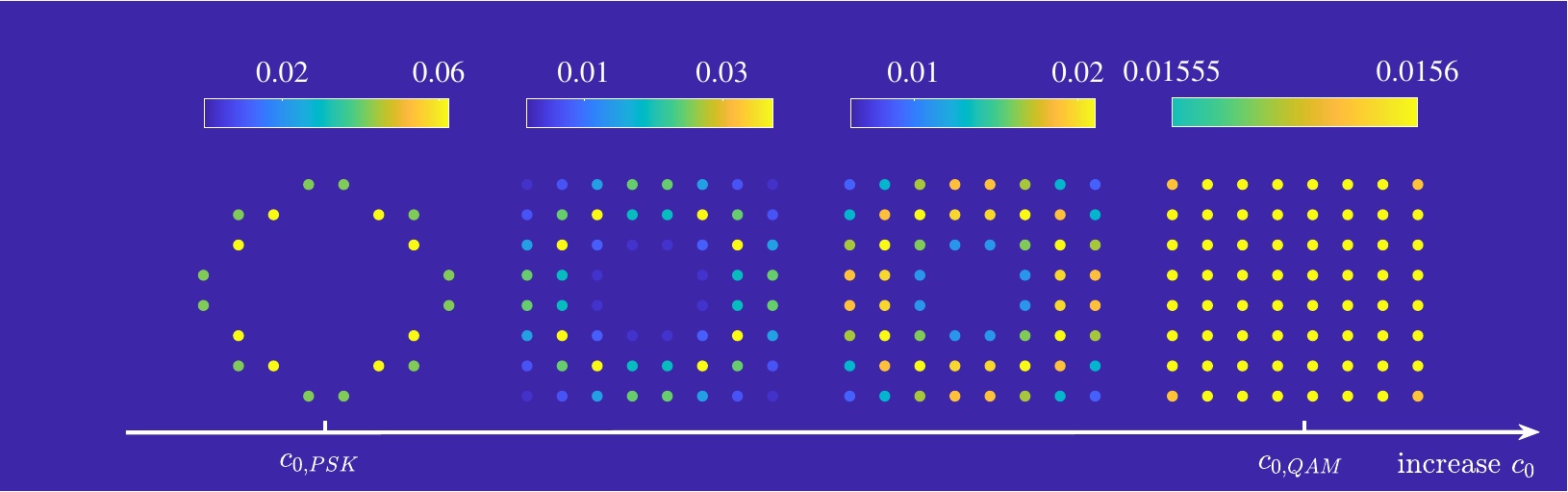}}
		\caption{64-QAM constellations with PCS, when $\text{SNR}_\text{in}=4$ dB.}\label{figure12}
	\end{figure}

	When $\text{SNR}_\text{in}=20$ dB in Fig. \ref{figure3c}, RF and WF achieve the similar sensing performance, as detailed in Remark 2. Moreover, their behaviors are evidently better than MF, since the randomness of discrete symbols plays a more dominated role in affecting the sensing capability, which can be significantly eliminated with element-wise division operation. Besides, when PCS is utilized toward the best sensing performance (i.e. constant modulus constellations), the range profiles of MF, RF and WF can all be enhanced. Notably, when $\text{SNR}_\text{in}\rightarrow+\infty$ (i.e. in a noise-free environment which is unrealistic), RF and WF can achieve extremely low sidelobes since the signaling randomness can be completely eliminated with element-wise division $\text{Var}(\chi)=0$, regardless of the constellation distribution. In contrast, $\text{Var}(\chi)>0$ still holds for MF as the optimized QAM is a pseudo PSK constellation, since its discrete symbols do not meet constraints of unit modulus and equally spaced phases simultaneously \cite{du2024reshaping}. This is also revealed in Fig. \ref{figure12}.
	
	Additionally, corresponding velocity profiles are depicted in Fig. \ref{figure4}, which are similar to range profiles in Fig. \ref{figure3}. 

	\begin{figure}[!t]
		\centering
		\includegraphics[width=0.95\linewidth]{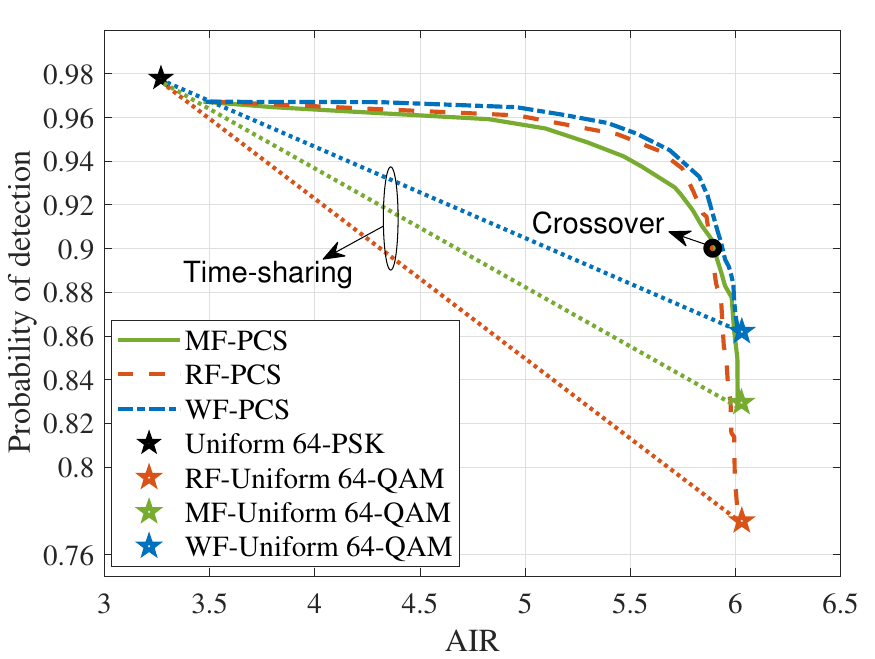}
		\caption{S\&C trade-off curves with MF/RF/WF, when $\text{SNR}_\text{in}=4$ dB.}
		\label{figure5}
	\end{figure}
	
	\subsection{Flexible S\&C Trade-off}
	Since the range-velocity profiles in Fig. \ref{figure3} and Fig. \ref{figure4} only illustrate the sensing performance of different filtering schemes in the context of uniform QAM constellation and the optimized PCS constellation, their capabilities of balancing between S\&C performance remain unsolved. To this end, we first depict 64-QAM constellations with PCS optimizations in Fig. \ref{figure12}, where MF, RF and WF are compared when $\text{SNR}_\text{in}=4$ dB. Evidently, with increased $c_0$ tailored for MF, RF and WF, respectively, corresponding constellations are all reshaped from uniform PSK toward uniform QAM. Notably, PCS with higher-order QAM (e.g. 64-QAM and 1024-QAM) cannot yield a constant modulus constellation, as mentioned previously.
	
	Next, to quantify the effect of PCS optimizations on S\&C performance, we explicitly characterize S\&C trade-off regions in Fig. \ref{figure5}. Specifically, we exploit the AIR as the communication metric with a tiny communication noise power, and the CFAR algorithm to detect a weak target nearby the strong target, where the scenario is similar to that in \cite{du2024reshaping}. Therefore, the probability of detection is used to quantify the sensing performance.
	Note that we sweep $c_0$ in (\ref{equ36}) to obtain the trade-off curves, as mentioned in Remark 6. Overall, the observations in Fig. \ref{figure5} are consistent with  those in Fig. \ref{figure3} and Fig. \ref{figure4}. Above all, MF and MMF schemes can achieve a flexible S\&C trade-off with PCS optimizations, which are superior to the time-sharing strategy \cite{xiong2023fundamental}. Additionally, WF can approach the outer boundary of S\&C in medium $\text{SNR}_\text{in}$ case over MF and RF. 
	Interestingly, RF and MF trade-off curves still have a crossover point in Fig. \ref{figure5} under non-uniform QAM constellations\footnote{This behavior can be explained with reference to Fig. \ref{figure10}. For instance, as $c_0$ is varied from uniform QAM to uniform PSK, the curves corresponding to RF and MF will gradually approach that of uniform PSK. Nevertheless, the crossover point persists, but the corresponding $\text{SNR}_\text{in}$ varies.}. Different from uniform QAM modulations as stated in Remark 2, when PCS is exploited, the input distribution $p(x)$ at the cross point in Fig. \ref{figure5} can be theoretically computed by solving 
	\begin{equation}
		\left\{
		\begin{aligned}
			\hat{p}(x)= & \arg\min_{p(x)} \left\vert\text{SNR}_\text{in}-\frac{\mathbb{E}\left\{ {|x|^{-2}} \right\}-1}{\mathbb{E}\left\{|x|^4\right\}-1}\right\vert^2
			\\
			s.t. \ 
			& C_2, C_3, C_4.
		\end{aligned}
		\right.
	\end{equation}

	\subsection{Measurement Validation}
\begin{figure}[!t]
		\centering
		\includegraphics[width=0.85\linewidth]{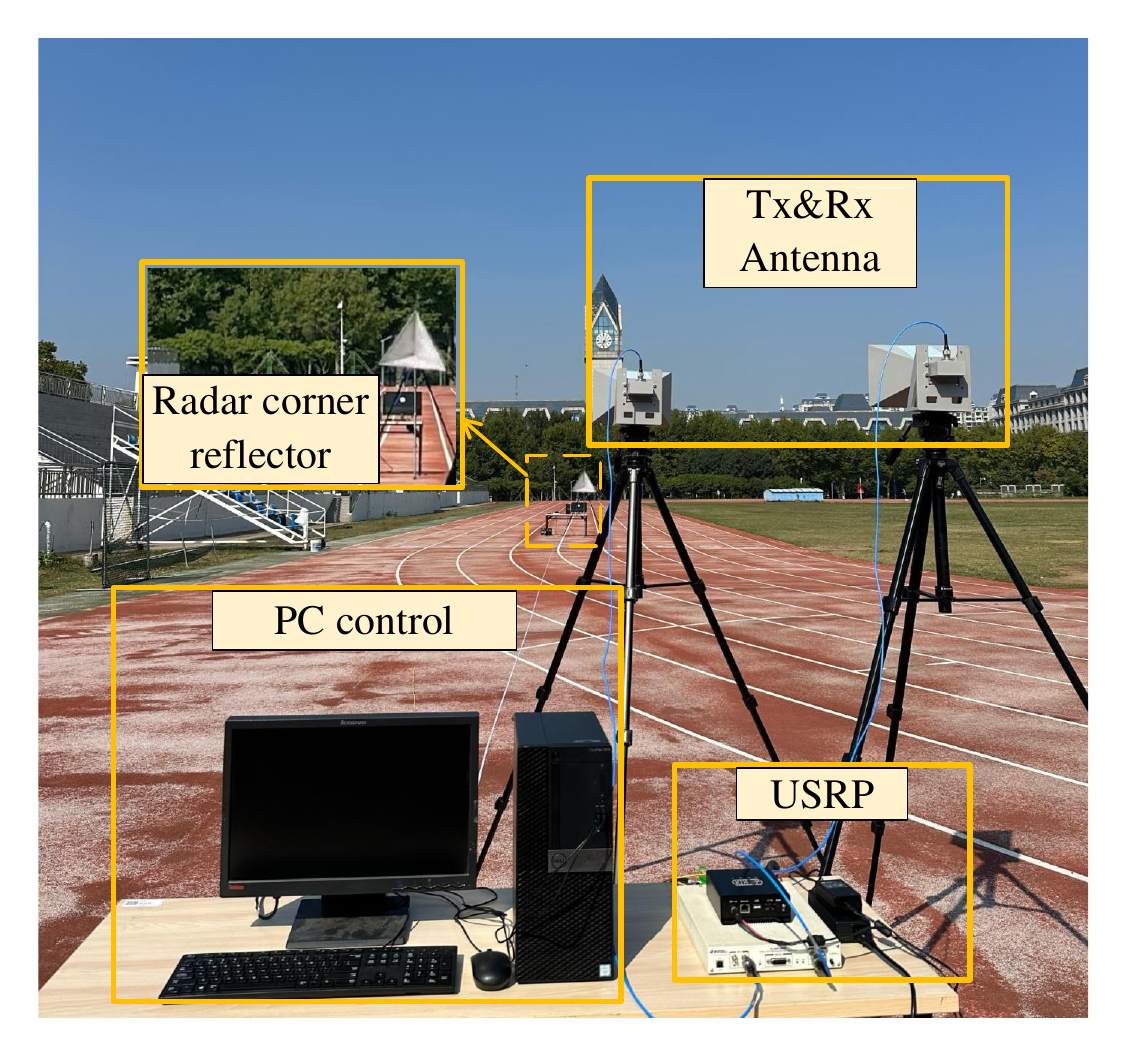}
		\caption{Scenario of field experiments and USRP prototype.}
		\label{figure6}
	\end{figure}

	To demonstrate the efficiency of proposed PCS approach in enhancing sensing behaviors, a field experimental system based on universal software radio peripheral (USRP) prototype \cite{wang2023waveform,liu2019implementation}, is depicted in Fig. \ref{figure6}, where a static corner reflector positioned in the radial distance of $15$ meters, is used as the interested radar target. Throughout experiments, USRP transmits the 64-QAM modulated OFDM ISAC signals with $100$ symbols with $1024$ subcarriers, where the transmit power is $20$ dBm, the antenna gain is $15$ dB, the carrier frequency is $3$ GHz, and the signaling bandwidth is $100$ MHz. All experimental results are averaged with $64$ individual observations.
		
\begin{figure*}[!t]
		\centering  
		\subfigure[MF]{
			\includegraphics[width=2.32in]{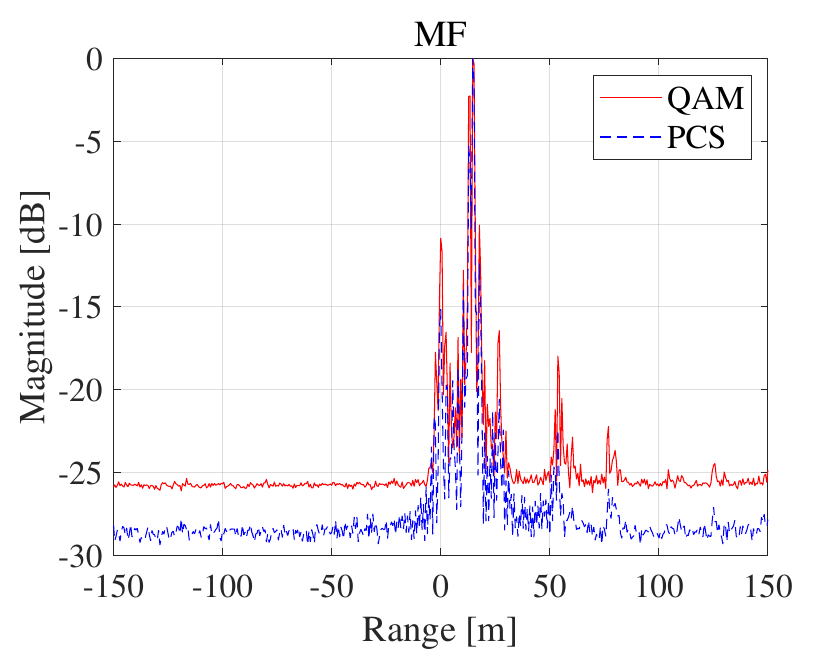}}
		\subfigure[RF]{
			\includegraphics[width=2.27in]{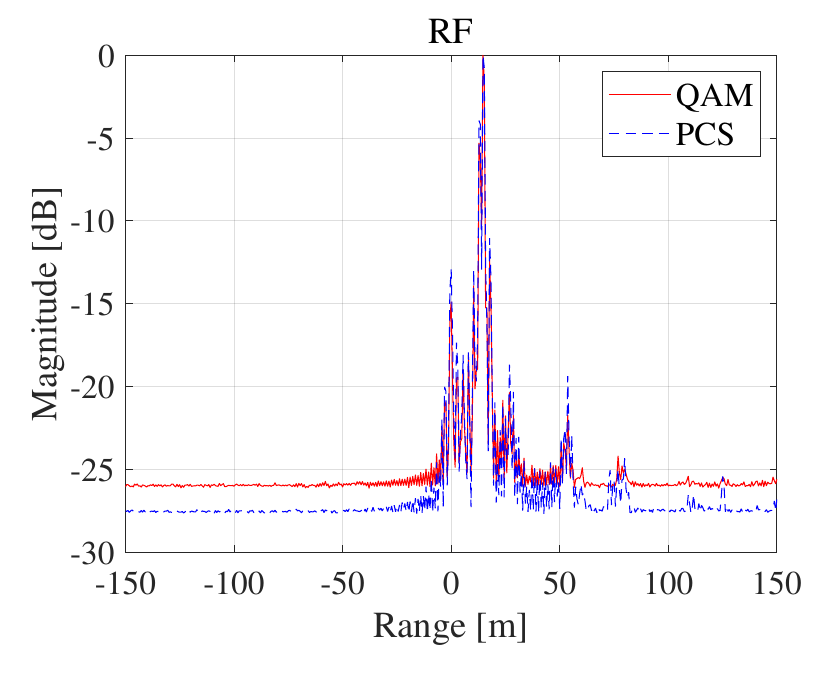}}
		\subfigure[WF]{
			\includegraphics[width=2.16in]{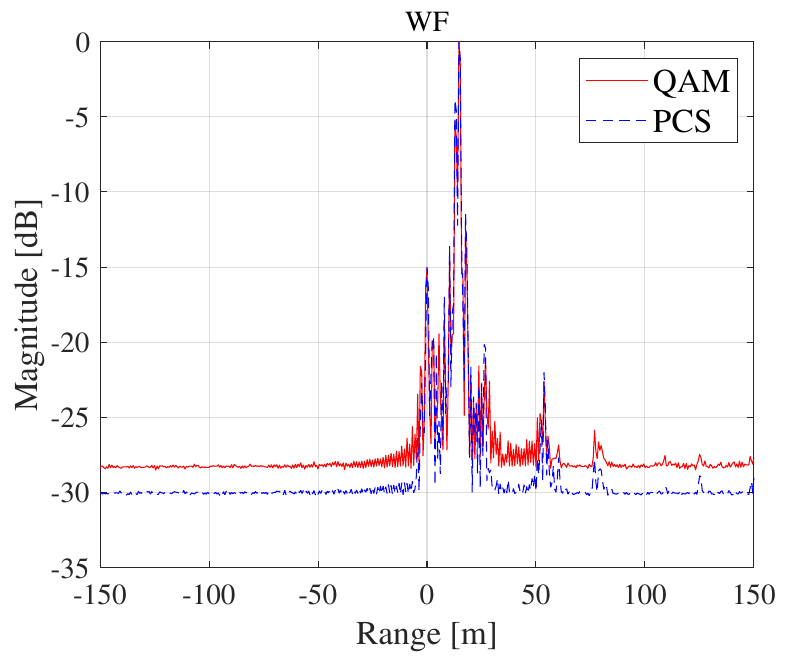}}
		\caption{Range profiles of MF/RF/WF with/without PCS optimizations: experimental results.}\label{figure7}
	\end{figure*}

\begin{figure*}[!t]
		\centering  
		\subfigure[MF]{
			\includegraphics[width=2.27in]{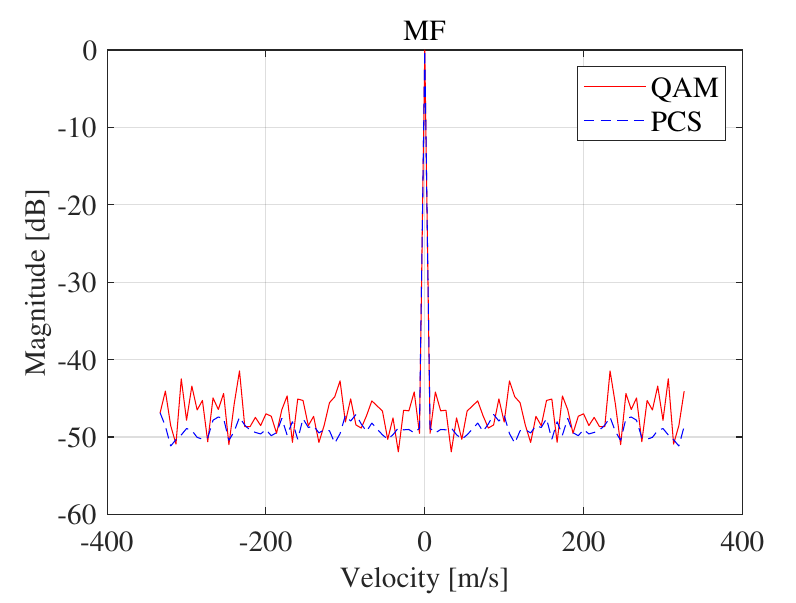}}
		\subfigure[RF]{
			\includegraphics[width=2.2in]{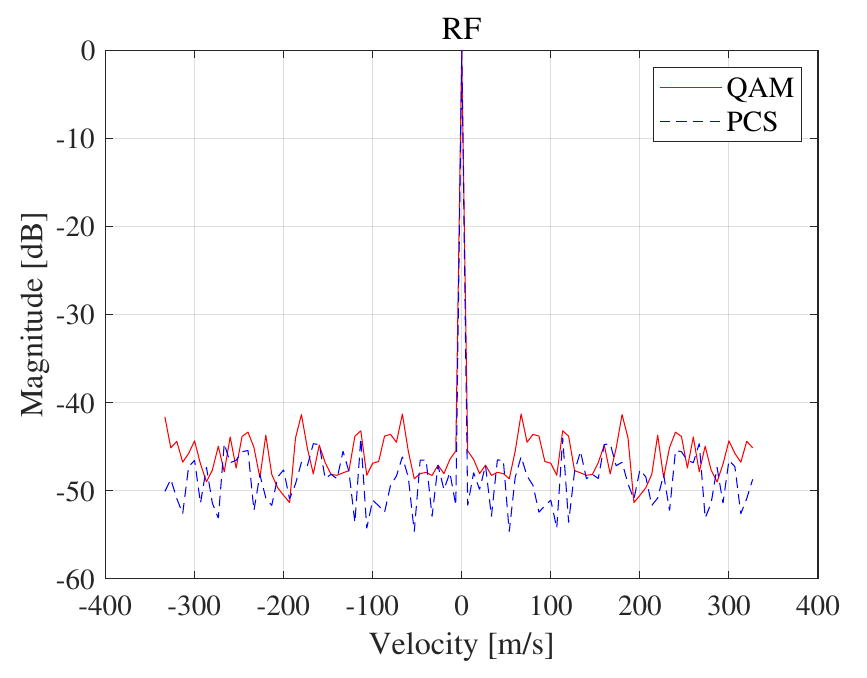}}
		\subfigure[WF]{
			\includegraphics[width=2.225in]{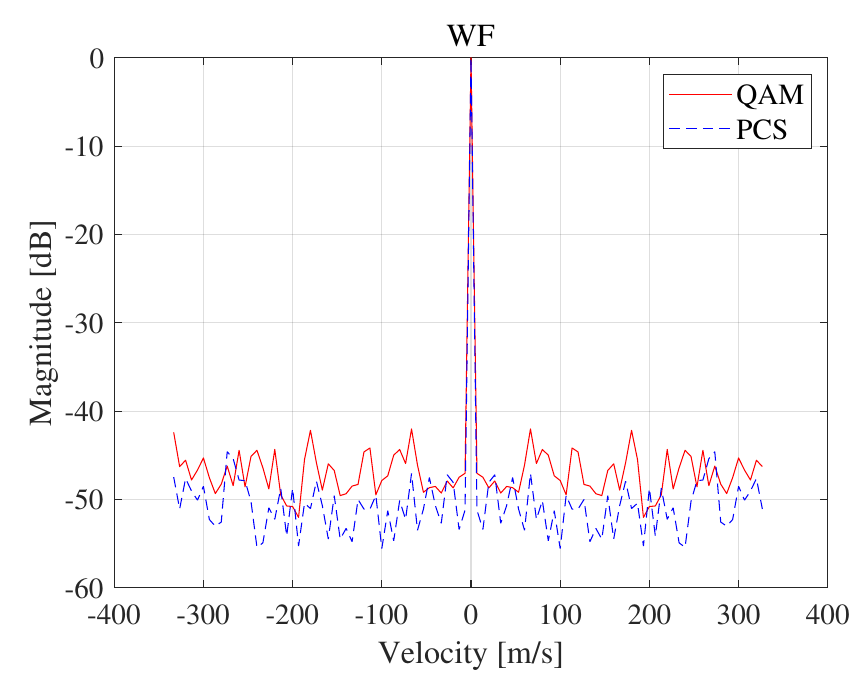}}
		\caption{Velocity profiles of MF/RF/WF with/without PCS optimizations: experimental results.}\label{figure8}
	\end{figure*}

	We carried out two groups of experiments with different filtering schemes: 1) ISAC signaling with uniform 64-QAM modulation, and 2) ISAC signaling with 64-QAM-PCS optimization toward the best sensing performance. After receiving and processing echoes with MF, RF and WF, the range and velocity profiles are shown in Fig. \ref{figure7} and Fig. \ref{figure8}, respectively. To be specific, the range profiles in Fig. \ref{figure7} exhibit many peaks, where the maximum peak (i.e. the corner reflector) is located at the distance of $15$ meters, and other peaks are contributed by clutter and self-interference. 
	Evidently, three filters can achieve an obvious sensing gain (2-3 dB) benefiting from PCS optimization, which can further improve the detection probability compared with the uniform QAM modulated OFDM ISAC signals. Moreover, our measurements demonstrate that WF achieves the highest sensing performance both with and without PCS. The velocity profiles in Fig. \ref{figure8} are extracted in accordance with the $15$ meters-range slices, which also verify the efficiency of PCS in enhancing Doppler performance. 
	
	Additionally, the optimized input distribution can be precomputed and stored in a codebook offline, then deployed online. The advantages clarified above can lay a solid foundation for practical ISAC applications in future 6G networks.

	\section{Conclusion}\label{sec6}	
	We propose a unified PCS approach to explore the trade-off between S\&C performance under OFDM signaling. Unlike the method in \cite{du2024reshaping}, which is limited to MF, our approach is compatible with both MF and MMF schemes. Specifically, we first derive the MSE of sensing CSI, the $\text{SNR}_\text{out}$, and the ISLR for MF, RF, and WF. We then establish the relationship among these three sensing criteria. By optimizing the input distribution of constellations, we maximize the AIR while constraining the MSE, the transmit power, and the probability simplex constraint. Utilizing a tailored MBA algorithm, we efficiently solve this optimization problem. Simulations provide fair comparisons among the three filtering schemes, demonstrating the PCS approach's capability to balance S\&C performance. Additionally, field experiments validate the effectiveness of our unified PCS approach and its potential for practical location-aware services in future 6G networks.

	
	\appendices
	
	\section{Proof of Corollary 1}\label{appendix2}
	In regard to (\ref{dft}), applying the well known Parseval’s theorem yields
	\begin{align}\label{eqxxx}
		\sum\nolimits_{n,m} \chi_{n,m}^2 = \sum\nolimits_{k,p} |r(k,p)|^2.
	\end{align}
	Then inserting (\ref{r00}) and (\ref{eqxxx}) into (\ref{eq33}) yields 
	\begin{equation}\label{islr2}
		\begin{aligned}			
			\text{ISLR} & = \frac{ \mathbb{E} \left\{\sum_{n,m} \left(\chi^2_{n,m}-\frac{2}{\sqrt{NM}}r(0,0)\chi_{n,m} + \frac{r^2(0,0)}{{NM}}\right) \right\} }{\mathbb{E} \left\{r^2(0,0)\right\}}
			\\ & = \frac{\mathbb{E} \left\{\sum_{k,p} |r(k,p)|^2-r^2(0,0)\right\}}{\mathbb{E} \left\{r^2(0,0)\right\}}.
		\end{aligned}
	\end{equation}
	The proof is thus completed.
	
	\section{Proof of Theorem 1}\label{appendix3}
	We denote $\frac{r(0,0)}{\sqrt{NM}}+\Delta = 1$, and thus formulate (\ref{eq34}) at the top of next page,
	\begin{figure*}[!t]
		\begin{equation}\label{eq34}
			\begin{aligned}
				\sum\nolimits_{n,m} \left(\chi_{n,m}-1\right)^2 
				& = \sum\nolimits_{k,p} |r(k,p)|^2 - 2\left({r(0,0)}/{\sqrt{NM}}+\Delta\right)r(0,0) + NM\left({r(0,0)}/{\sqrt{NM}}+\Delta\right)^2 \\
				& = \sum\nolimits_{k,p} |r(k,p)|^2 - r^2(0,0) + NM\left(1-{r(0,0)}/{\sqrt{NM}}\right)^2
			\end{aligned}
		\end{equation}
		\rule{18cm}{1.0pt}
	\end{figure*}	
	where the sum of the first two terms and the third term in the last equation of (\ref{eq34}) represent the numerator of ISLR (i.e., the energy of sidelobes) and the peak energy loss due to MMF, respectively. Here, the peak energy loss is referred to $\mathbb{E}\left\{
	\frac{r(0,0)}{\sqrt{NM}}\right\}\leq1$ for MMF.
	Next, multiplying the left of (\ref{theorem}) by $\mathbb{E} \left\{r^2(0,0)\right\}$ yields (\ref{eq26}) at the top of next page.
	\begin{figure*}[!t]
		\begin{equation}\label{eq26}
			\begin{aligned}
				\sigma^2_{\alpha} \bigg(\mathbb{E} & \bigg\{\underbrace{\sum\nolimits_{k,p} |r(k,p)|^2 - r^2(0,0) + NM\left(1-\frac{r(0,0)}{\sqrt{NM}}\right)^2}_{(\ref{eq34})}\bigg\} \bigg)   +\sigma^2\sum\nolimits_{n,m}\mathbb{E}\left\{|g_{n,m}|^2 \right\}
				\\ & = \sigma^2_{\alpha} \mathbb{E} \left\{\sum\nolimits_{n,m}\left( \chi_{n,m}-1\right)^2\right\} + \sigma^2\mathbb{E} \left\{\sum\nolimits_{n,m}|g_{n,m}|^2\right\}
				= \mathbb{E}\{\varepsilon^2\}
			\end{aligned}
		\end{equation}
		\rule{18cm}{1.0pt}
	\end{figure*}	
	This completes the proof of Theorem 1.

	\section{Proof of Theorem 2}\label{appendix4}	
	In accordance with the definition of $\mathbb{E}\{\varepsilon^2\}$ in (\ref{mse}), one may reformulate 
	\begin{equation}\label{mse3}
		\begin{aligned}
			& \mathbb{E}\{\varepsilon^2\} = \mathbb{E}\left\{\left\Vert {\hat{\bf{H}}-\bf{H}} \right\Vert^2_F \right\}
			= \mathbb{E}\left\{\left\Vert \bf{F}^H_N\left(\hat{\bf{H}}-\bf{H}\right)\bf{F}_M \right\Vert^2_F \right\}
			\\ & = \left\Vert \mathbb{E}\left\{\bf{F}^H_N\left(\hat{\bf{H}}-\bf{H}\right)\bf{F}_M \right\} \right\Vert^2_F + \text{Var}\left(\bf{F}^H_N\left(\hat{\bf{H}}-\bf{H}\right)\bf{F}_M\right),
		\end{aligned}
	\end{equation}
	where
	\begin{align}
		\left\Vert \mathbb{E}\left\{\bf{F}^H_N\left(\hat{\bf{H}}-\bf{H}\right)\bf{F}_M \right\} \right\Vert^2_F = 0 
	\end{align}
	can be readily verified. The proof is thus completed.

\ifCLASSOPTIONcaptionsoff
\newpage
\fi

\bibliographystyle{IEEEtran}
\bibliography{reference}

\end{document}